**Development and Ground Demonstration of a Modular Astrobiological Experiments (MAEx) payload for autonomous biological monitoring in Low Earth Orbit (LEO)**

**Authors:**

Yamini Jangir[1*], Vinay Nayaka[1], Abhishek M. Kulkarni[1], Samrat Ghosh[1], Mubashir Ali[1], Dharshan Hegde[2], Saurav Mishra[1], Kunal P. Mooley[1], Sujata Malik[3], Shivani Vinod[4], Kalpesh Wagh[1], Hariharan V.C.[3], Arunima Saha[5], Amey Bagare[1], Mukuljeet S. Mehrolia[1], Saravanan Matheswaran[3], Ashwani K. Thakur[3], Suvendu Samanta[6]

*Corresponding author (jangir@iitk.ac.in)

**Affiliation:**

[1]Department of Space, Planetary & Astronomical Sciences & Engineering, Indian Institute of Technology Kanpur, Kanpur, UP 208016, India.

[2]Department of Aerospace Engineering, Indian Institute of Technology Kanpur, Kanpur, UP 208016, India.

[3]Department of Biological Sciences and Bioengineering, Indian Institute of Technology Kanpur, Kanpur, UP 208016, India.

[4]Department of Biotechnology, Ramaiah Institute of Technology, Bangalore, KA-560054, India

[5]Department of Earth, Atmospheric, and Planetary Sciences, Purdue University, West Lafayette, IN 47907, USA

[6]Department of Electrical Engineering, Indian Institute of Technology Kanpur, Kanpur, UP 208016, India.

**Contributions:**

Payload development and characterization: SG contributed to payload development and initial characterization of the spectroscopy module. VN designed and functional characterized the imaging and electrochemical modules. MA contributed to payload design and downstream thermal validation. DH contributed to payload design and downstream vibrational analysis.

MSM supervised the initial circuit design of the potentiostat. SM (Saurav Mishra) performed image and spectroscopy analysis. Printed circuit board (PCB) fabrication was facilitated by SS.

Biological experiments: AB conducted the initial electrochemical lab-based experiments. SV and KW performed Shewanella growth and functional characterization of the potentiosat design. SM (Sujata Malik) performed protein aggregation experiments. HVC conducted the initial fungal growth experiments, and SM (Saravanan Matheswaran) provided the *Ustilago maydis* FB1 isolate. AKT provided γD-crystallin aggregates and BSA protein.

Project supervision and manuscript preparation: KM contributed to payload conceptualization and supervised the overall project. YJ conceived and supervised the project, coordinated the biological experiments, and drafted the original manuscript. YJ, SG, VN, MA, SM (Saurav Misra), KPM, and AKT contributed to manuscript review and editing. All authors reviewed and approved the final manuscript.

**Acknowledgments:**

AKT sincerely thanks Martin Zanni (Department of Chemistry, University of Wisconsin-Madison, USA) for providing the plasmid containing the γD-crystallin gene, which enabled overexpression of the γD-crystallin protein. AKT acknowledges financial support from IIT Kanpur under project grants IITK/BSBE/20100293. KPM thanks IIT Kanpur for their financial support through under project grants IITK/SPASE/2023222 and IITK-SPAS-2026384. YJ and KPM would like to thank R. Vijaya for facilitating UV LED characterization. YJ thanks the Advanced Imaging Center, IIT Kanpur, for providing access to its microbiology laboratory facilities. AS sincerely thanks Mouchandra Paul, Gouri S. Nair, and Kavikumar A. K. for training on microbiological techniques and spectrofluorometer facility (S. Ganesh, IIT Kanpur). KPM and YJ are grateful to Aloke Kumar for initial discussions.

**Ethics declarations:**

The authors declare no competing interests.

**Abstract**

The spaceflight environment presents unique physicochemical conditions for biological systems, including microgravity, ionizing radiation, altered fluid transport, and confined habitats. Investigating and harnessing biological processes under these conditions are essential to enable future space biomanufacturing approaches, including scalable biological production systems and bioregenerative life-support technologies for long-duration human exploration. However, translating any biological processes into fully operational space technologies requires building a compact, autonomous, and flight-compatible experimental platforms capable of investigating biological responses to various space stressor using multiple analytical technique. Here, we present the ground-based validation of Modular Astrobiology Experiment (MAEx) payload, a compact 3U platform designed to support future autonomous biological characterization in Low Earth orbit (LEO). MAEx integrates three analytical modules, 1) camera-based imaging, 2) absorption spectroscopy, and 3) microbial electrochemistry. To demonstrate platform versatility, the payload incorporates three biological systems: 1) electroactive bacterium *Shewanella oneidensis* MR-1, 2) the radiation-resistant fungus *Ustilago maydis* FB1, and 3) the human eye lens protein γD-crystallin. The imaging module is validated by temporally monitoring the growth of *S. oneidensis* MR1 and *U. maydis* FB1. The spectroscopy module is evaluated using all three biological systems to investigate microbial growth and protein aggregation. The electrochemical module was functionally characterized by temporal anodic current production associated with electrochemical growth of *S. oneidensis* MR1 population. Taken together, we independently demonstrate the functionality of the three analytical modules and provide a ground-based validation of MAEx payload as a TRL-4 platform. This payload provides a foundation for biomanufacturing related investigations integral to develop bioregenerative life-support systems, *in situ* resource utilization, and autonomous biological production in space. While environmental qualification for independent spaceflight remains ongoing, the ground-validated MAEx architecture is suitable for future integration into existing or planned orbital laboratories.

## 1. Introduction

Space imposes a range of physicochemical constraints on life, including altered gravity, ionizing radiation, and extreme temperature and pressure regimes, while spaceflight introduces additional engineered conditions such as microgravity and controlled atmospheres that shape biological responses[1–3]. Sustaining a long-term human presence beyond Earth requires minimizing dependence on supplies transported from Earth and enabling efficient recycling and utilization of available resources[4,5]. Biological systems offer an important approach towards addressing these challenges through *in situ* space biomanufacturing, which encompasses the harnessing microorganisms[6,7] and plants[8–11]. Other biological systems, including proteins, have revolutionized regenerative medicine in space[12,13]. More recently, an emerging direction in space biomanufacturing is the development of "made in space, used on Earth" approaches, in which products such as pharmaceuticals will be manufactured in the space for subsequent application on Earth[14]. While investigating individual biological systems is an important first step towards space biomanufacturing, long-term *in situ* biomanufacturing will require an integration of multiple complementary biological systems to function together.

Bioregenerative life-support systems (BLSS), are a proposed set of complex network of interconnected biological processes useful for wastewater recycling and regenerate essential resources such as food, water and oxygen, in space[15–17]. The European Space Agency's Micro-Ecological Life Support System Alternative (MELiSSA) program is a prominent example, developing a compartmentalized microbial and plant-based ecosystem for resource recovery and regeneration during long-duration space missions[18]. By emphasizing metabolic functionality and interdependence rather than individual taxa, this approach offers a blueprint for designing reduced yet resilient set of biological systems. Such systems, which are likely to operate within engineered enclosures in the orbit[19] or stationed on planetary surfaces, such as Moon[20] or Mars[21], must be optimized for bioresource production efficiency under space-relevant stressors. Optimizing a biological system for bioresource production first requires understanding its physiological responses to space-relevant environmental stressors.

Although terrestrial simulation platforms provide an important first step for investigating biological responses to microgravity[22,23] and radiation[24–26], the design and deployment of spaceflight payloads remain essential to accurately capture the full spectrum of space stressors experienced during spaceflight. Ground based analogs such as clinostats, rotating wall vessels, and random positioning machines attempt to average the gravity vector over time; however, they do not eliminate gravity itself and therefore cannot fully reproduce true microgravity conditions[27]. Residual accelerations, shear forces, and rotational artifacts introduced by these systems can alter fluid dynamics, mass transfer, and cell surface interactions in ways that are not true representative of spaceflight. These limitations have led to discrepancies in microbial growth, gene expression, biofilm formation, and metabolic activity when compared to observations from orbital platforms[28]. In addition, terrestrial systems often fail to capture the combined effects of microgravity with other space stressors such as chronic low dose radiation and operational perturbation. Hence, spaceflight payloads provide a critical validation platform, enabling direct measurement of biological processes under spaceflight conditions and supporting the development of robust life support systems for long duration human missions.

Several independent small-satellite missions have demonstrated the feasibility of conducting autonomous biological experiments in orbit, including GeneSat-1[29], PharmaSat[30], O/OREOS[31], EcAMSat[32], BioSentinel[33], and LabSat[34]. Recent work proposed an autonomous nanosatellite designed as a miniature research station for biological missions and compared it in detail with these earlier missions[34]. That comparison highlights that the missions used miniaturized optical sensing systems, including colorimetry and spectroscopy, to monitor biological responses to microgravity or radiation exposure. Of these missions, BioSentinel has been most successful. Launched aboard NASA's Artemis I in November 2022, BioSentinel, a 6U CubeSat, is till date the first free flying biological experiment in deep space, incorporating the metabolic indicator resazurin (alamarBlue) to assess physiological state of *Saccharomyces cerevisiae*[35]. The scientific objective was to investigate the effects of deep-space radiation on *Saccharomyces cerevisiae*; however, prolonged prelaunch storage compromised cell viability, resulting in inconclusive biological results. However, the technical design of BioSentnel will now inform the development of Lunar Explorer Instrument for Space Biology Applications (LEIA), a biological payload to be hosted on the Lunar surface[36]. Collectively, these missions highlight ongoing efforts to develop improved spaceflight instruments that meet biological experimental requirements and enable autonomous microbial investigations and monitoring of microbial responses to the space environment.

Analytical capabilities of small biological payloads have expanded beyond optical measurements. For example, electrochemical ion-selective sensors have been integrated with optical absorbance measurements in CubeSat-based platforms to monitor parameters such as pH, sodium ion concentration, and changes associated with microbial growth in fluidic wells[37].

These developments represent an important step towards biological monitoring by combining complementary sensing approaches with autonomous data acquisition. A further progression in this direction is the development of autonomous platforms capable of monitoring entire microecosystems within controlled, closed habitats. Such systems enable investigations of interactions amongst different biological systems and the biomonitoring at the ecosystem-level, including primary production, biomass dynamics, nutrient cycling, and organic matter decomposition. An example is a recently developed miniaturized bioregenerative microecosystem comprising a mixed community of photoautotrophic mosses (*Brachythecium* sp. and *Tortula* sp.) and heterotrophic soil-dwelling microarthropods, including *Folsomia candida*, *Sinella* spp., and members of the family *Euphthiracaridae*[38]. Such integrated systems provide opportunities to investigate the functioning and stability of multispecies biological communities under controlled environmental conditions

Despite these advances, the integration of complementary analytical techniques for autonomous monitoring of multiple biological systems and processes within a single compact spaceflight platform remains missing. Existing missions have largely focused on monitoring specific biological responses using individual or limited combinations of sensing approaches. Addressing this gap, we present the architectural design and ground validation of a Modular Astrobiology Experiment (MAEx) payload, a 3U platform (10 x 10 x 30 $cm^3$) designed to investigate microbial growth (via imaging and absorption spectroscopy), metabolic activity (via fluorescence spectroscopy and electrochemical signal), and protein aggregation (via absorption spectroscopy), providing multiple lines of evidence for biological function through different levels of organization. Functional ground-validation, of MAEx, employed three biological models with relevance to future space biomanufacturing and bioregenerative life-support systems: 1) the dissimilatory metal-reducing bacteria (DMRB), *Shewanella oneidensis* MR-1[39], which has potential applications in wastewater treatment, bioelectrochemical energy generation and biomining; 2) the UV-radiation resistant fungus, *Ustilago maydis* FB1[40], which can serve as a platform for producing bioactive compounds, including melanin; and 3) human γD-crystallin[15,16], a major structural protein of the eye lens whose aggregation leads to cataract disease. Together, this work provided the architectural design and establishes a ground-validated experimental platform for investigating biological processes relevant to future space-based resource conversion, biological production, and biomolecular applications.

## 2. MAEx Science Goals

The design and validation of the Modular Astrobiology Experiment (MAEx) platform incorporates biological systems spanning bacterial, eukaryotic, and protein-based model systems to capture diverse aspects of biological function under space-relevant conditions. These systems were selected to investigate microbial metabolism, stress adaptation, and macromolecular assembly dynamics.

*Shewanella oneidensis* MR-1 is a facultative anaerobic, psychrotolerant bacterium capable of respiring a wide range of electron acceptors, including oxygen and insoluble metal oxides[41]. A defining feature of this organism is its ability to perform extracellular electron transfer (EET)[42], enabling reduction of minerals and interaction with solid-phase electron acceptors. This metabolic versatility underpins its relevance to bioelectrochemical systems, electromicrobial production, and biomining applications[43]. The capacity of *S. oneidensis* to couple metabolism with electron transfer makes it a suitable model for studying energy flux under space-relevant constraints. Previous studies have demonstrated its ability to reduce iron oxides in simulated Martian regolith (JSC-Mars1), resulting in significant iron mobilization[44], highlighting its potential role in *in situ resource utilization* (ISRU) strategies, including metal recovery and redox-driven processes in extraterrestrial environments. Investigating the growth physiology and electron transfer dynamics of *S. oneidensis* under spaceflight conditions is therefore relevant for sustainable space exploration.

*Ustilago maydis* is a eukaryotic fungal model organism with potential for biotechnological products, including itaconic acid[45]. Moreover, *U. maydis* produces melanin upon UV and ionizing radiation exposure, via DNA repair mechanisms[46], making it relevant to investigations of fungal stress tolerance and biological melanin production[40]. Melanized infused polylactic acid (PLA), a common 3D printing filament, lost 89% less mass compared to pure PLA when exposed to harsh LEO environment suggesting biobased architecture could support sustainable, long term space exploration[47]. Melanized clones of fungal taxa, including *Cryptococcus neoformons*[48], *Cladosporium sphaerospermum*[49] and *Aspergillus niger*[50], also show higher protection under radiation exposure, Moreover, mycelium based composites (MBCs) combining fungal hyphae with feedstock offer potential of radiation resistant structural composites through mycotecture[51]. *U. maydis* therefore provides an opportunity to extend these investigations using a fungus that combines melanin production with metabolic versatility, alternative feedstock utilization and an established capacity for producing industrially relevant compounds.

To complement biomolecular models, MAEx incorporates human γD-crystallin (HγD-Crys) as a model system to investigate protein stability and aggregation under microgravity. A human eye lens is composed of ~90% crystallin proteins, encompassing the α-, β-, and γ-crystallin families[52]. The nuclear region cells of the lens are densely packed with crystallins and lack the capacity for protein synthesis or turnover, resulting in exceptionally long-lived proteins. Consequently, cumulative post-translational modifications, oxidative stress and radiation exposure can destabilize crystallin proteins, promoting their aggregation and ultimately contributing to cataract formation[53]. On Earth, highly conserved pairs of tryptophan residues hosted in human γD-crystallin have been proposed to act as a protection layer against UV-induced photodamage[54]. Further similar investigation, under spaceflight conditions, will inform an astronauts' ocular health, particularly given the concern over radiation-associated cataract development during long term space missions[55–58]. Previous investigations have also demonstrated that altered gravity can influence protein aggregation and crystallization. An experiment aboard the ISS using a ring-sheared-drop reactor characterized insulin fibrillization under interfacial shear, demonstrating seeded fibril formation and concentration-dependent gelatin[59]. In addition, ISS experiments with the therapeutic monoclonal antibody pembrolizumab produced more uniform crystalline suspensions than ground controls, providing insights into the development of injectable protein formulations [60]. These studies highlight the importance of investigating protein aggregation kinetics under both altered gravity and radiation, with implications on ocular health and biopharmaceutical development.

Collectively, these biological systems provide complementary models to functionally validate the three analytical modules: imaging, optical spectroscopy, and electrochemical capabilities. They represent processes relevant to future space biomanufacturing, including microbial resource conversion, fungal metabolite production, and protein aggregation. Although this study focuses on ground-based development and validation, the modular architecture of MAEx provides a foundation for extending these investigations to additional biological systems under spaceflight conditions.

### 3. Payload Description

The MAEx payload is designed as a modular bioanalytical system comprising of six primary subsystems: a power distribution module, a control and data handling module, three biological analytical modules (imaging, spectroscopy, and electrochemistry), and a housekeeping module (Figure 1). This architecture enables coordinated acquisition of biological and environmental data under spaceflight-relevant conditions. The detailed components of the payload are listed (Supplementary Table 1).

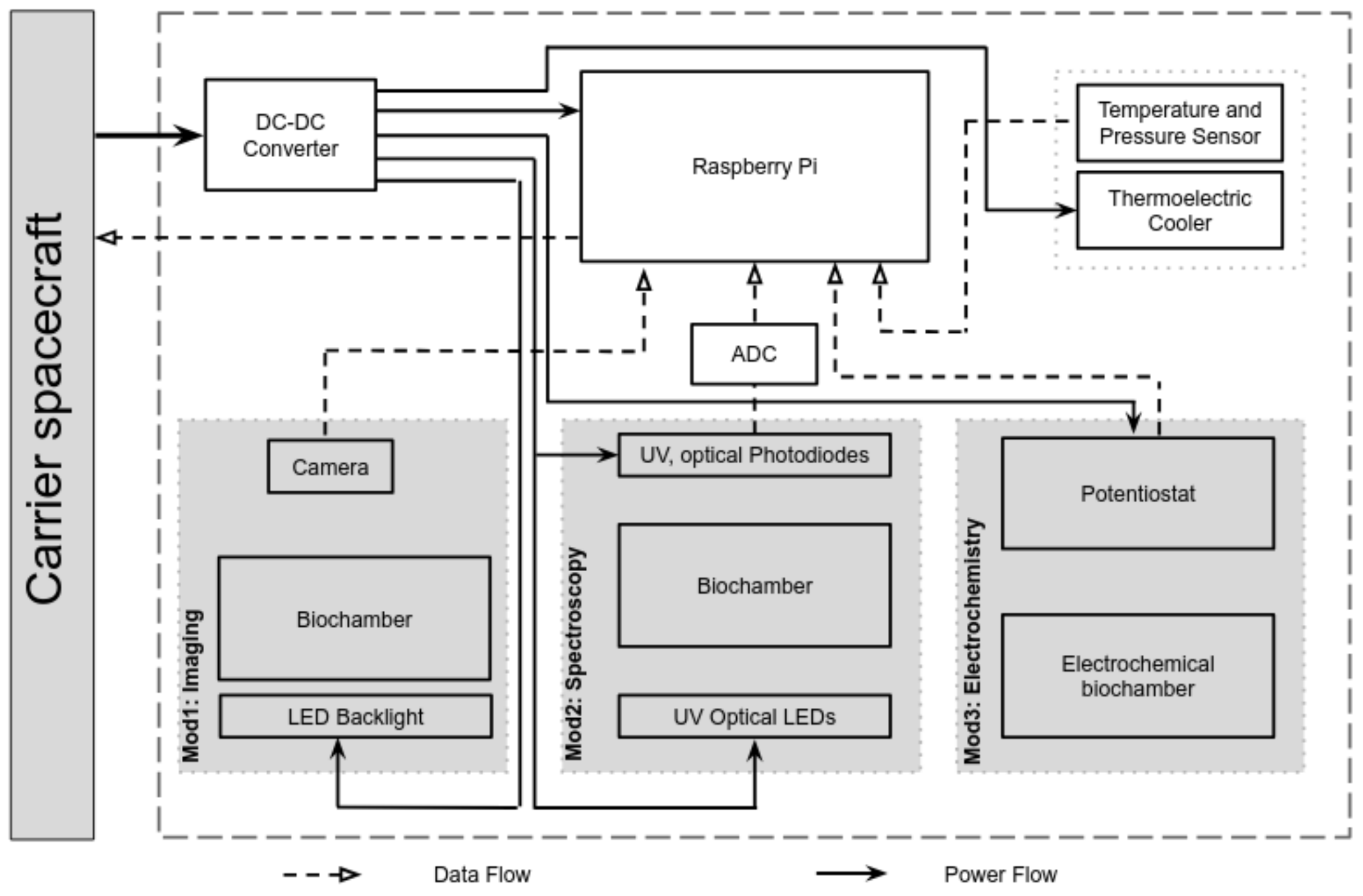

**Figure 1:** Power and signal flow diagram. Individual components are represented in blocks. The grey blocks with dotted lines as boundaries are three different analytical modules. The power flow is indicated with the solid arrows which flow to each component from the DC-DC converter which takes power from the carrier platform. The dashed arrows show the signal/data path which flows to RPi from every experiment and sensor and gets stored in the RPi. In a suitable time interval, the stored data is transferred to the carrier platform to be relayed to the ground station.

### 3.1 Payload Architecture

The MAEx payload is designed to satisfy spaceflight constraints associated with volume (3U; 10 × 10 × 30 $cm^3$), mass (< 3 kg), and power consumption (~6.5 W idle and <20 W peak using a 24-36 V power bus), while maintaining compatibility with autonomous biological experimentation. Analytical operations are implemented, using a Raspberry Pi 3 Model B+, with experimental data stored locally onboard and subsequently transmitted through the parent satellite communication system. The payload additionally incorporates regulated power distribution, periodic data logging, and timestamp synchronization through an onboard real-time clock (RTC) module.

**Figure 2:** Structural overview of the Modular Astrobiology Experiment (MAEx) payload and integrated analytical modules. (A) Fully assembled MAEx payload showing the vertically integrated imaging, spectroscopic, and electrochemical modules within a 3U CubeSat-compatible architecture (10 × 10 × 30 $cm^3$). (B) Exploded view of the payload architecture illustrating modular subsystem integration (outer casing not shown). Electronic components and control systems are highlighted in green, custom LED and photodiode printed circuit boards (PCBs) are shown in blue, biochambers and electrochemical biochambers are shown in olive-grey, and the structural electronics cage and supporting framework are rendered in metallic silver-grey tones. The figure additionally highlights integration of the Raspberry Pi 3 Model B+controller, in-house potentiostat modules, (putative) thermoelectric cooling system, and a temperature-pressure sensor within the payload structure.

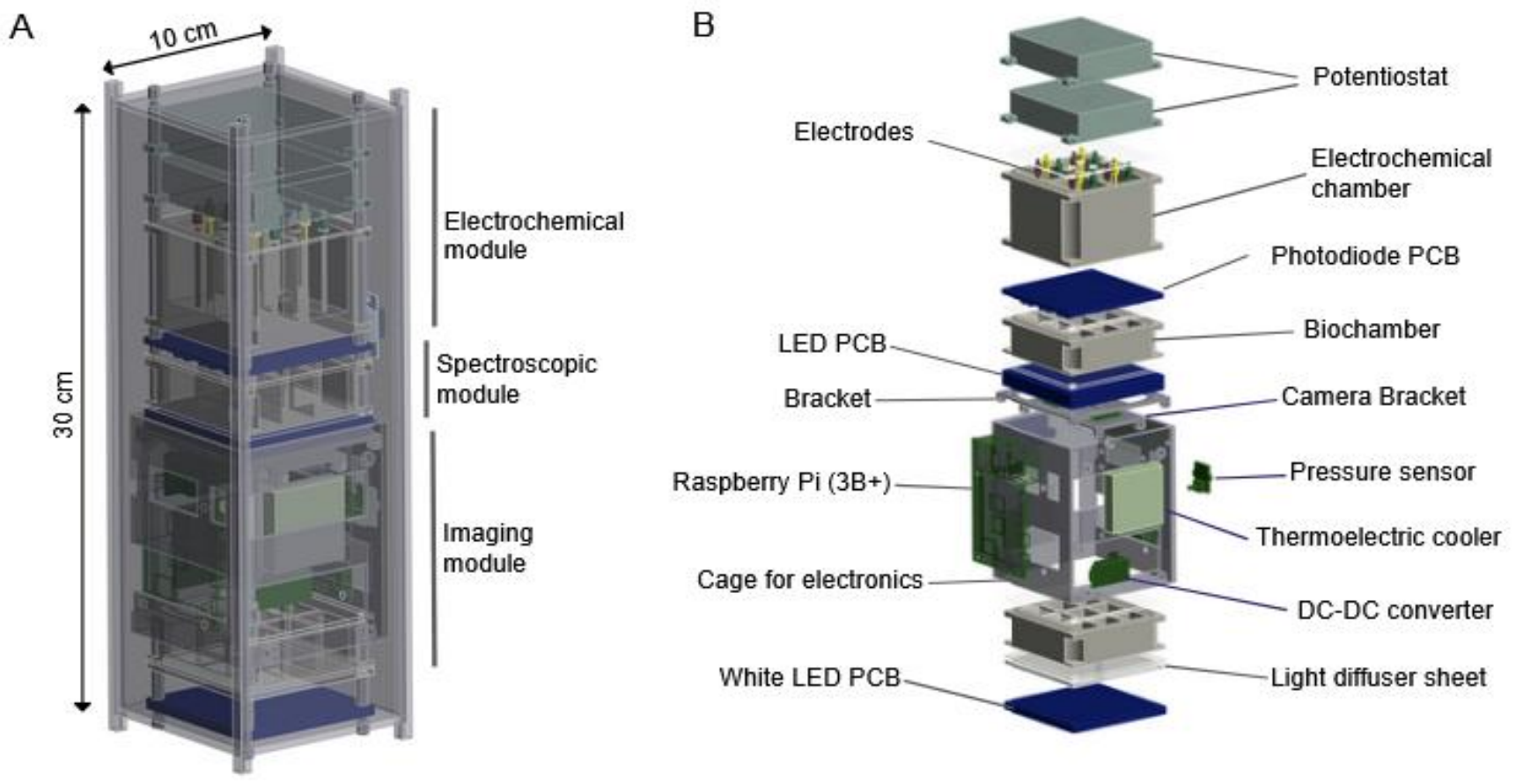

The primary structural framework is fabricated from Al 6061 alloy, while the biological and electrochemical chambers are constructed from polycarbonate. MAEx follows a fully modular, non-welded structural design in which all components are assembled using threaded rods, fasteners, and interlocking structural elements. The three experimental modules are vertically stacked using spacers for mechanical separation and subsystem protection, while electronic boards are mounted within a dedicated cage structure secured to the payload walls. These strategies are employed for optimizing structural rigidity and vibration tolerance during launch conditions.

The Raspberry Pi 3 B+ serves as the central processing and control unit for payload operation and was selected because of its low power consumption and compatibility with Python-based autonomous experiment control. Currently, the custom printed circuit board deployed on the payload uses FR-4 printed circuit boards (PCBs). However, future spaceflight deployment will host space-grade, MIL-specification laminate materials, Garolite G-10/FR-4, to meet the required mechanical, electrical-insulation, and flame-retardancy specifications.

### 3.2 Analytical Modules

The MAEx payload integrates three complementary analytical modules: 1) imaging, 2) spectroscopy, and 3) electrochemistry, to enable characterization of biological systems. These modules will monitor microbial growth and protein aggregation across spatial; and temporal scale.

To investigate microbial growth visually, we included the imaging module, using a Raspberry Pi Camera Module 3 Wide for ground-based testing (Figure 2B). Each culture well is positioned above an LED illumination array (XL-5050UWC, XINGLIGHT). Images are acquired at predefined intervals (0.5 h) and stored in PNG format with timestamp-based filenames. Two cameras with overlapping fields of view will be deployed, each covering approximately two-thirds of the biochamber assembly. This configuration enables reliable visualization of microbial growth dynamics and spatial heterogeneity during long-duration experiments. Although Raspberry Pi-based imaging systems have demonstrated utility in automated time-lapse imaging applications, including plant phenotyping and long-duration biological monitoring[61], their suitability of space flights application, particularly their tolerance of ionizing radiation, requires further validation. Alternate imaging systems for spaceflight MAEX could include, previously flown spaceflight-compatible Ascent A8050[62] (Apogee Imaging Systems, California, USA) used for fluorescence imaging of *Arabidopsis* specimens under controlled conditions, while CMOS camera ELP-USB8MP02G[34] (Shenzhen Ailipu Technology, China) for imaging fungi from the *Fusarium culmorum* family and seeds of *Lepidium sativum.* For ground-based functional validation of the imaging module, we employed Raspberry Pi Camera to monitor freshly inoculated biochambers containing *S. oneidensis* MR1 and *U. maydis* FB1.

Complementing the imaging module, the spectroscopy module quantifies biological growth and biochemical changes by measuring optical transmission through each culture wells (Figure 2B). Previous studies have demonstrated the development of compact LED-photodiode spectroscopy systems for autonomous biological monitoring across terrestrial and spaceflight environments. Terrestrial platforms such as MicrobeMeter (optical attenuation at 601 nm)[63], TubeOD (optical attenuation at 591 nm)[64], and ClampOD (optical attenuation at 591 nm)[64] employed low-cost LEDs, silicon photodiodes, and microcontroller-based electronics for continuous optical density measurements in microbial cultures, anaerobic systems, and thermophilic growth environments, respectively. Turbidimeter (optical attenuation at 598 nm)[65] was adapted to monitor microbial growth in blood culture using integrated 3D-printed optical assemblies. In spaceflight applications, missions including GeneSat-1 (470 nm for Ex/Em; optical attenuation at green nm)[59], PharmaSat (470, 535, and 615 nm)[60], O/OREOS (470, 525 and 615 nm)[61], EcAMSat (470, 525 and 615 nm)[62], and BioSentinel (570, 630, and 850 nm)[63] employed multiwavelength LED-photodiode optical systems for autonomous monitoring of microbial growth under orbital conditions. Building on these earlier systems, the MAEx spectroscopy module incorporates ultraviolet (UV; 365 nm; MT3650W3-UV, Marktech Optoelectronics) and red (630 nm; generic) LEDs as light sources, paired with corresponding photodiodes for signal detection in the UV (MICROFJ-30035-TSV-TR, onsemi) and visible (OPT101P, Texas Instruments) ranges, to be referred as UV and red channel henceforth. The red channel is used to measure an apparent absorbance signal (turbidity) for monitoring microbial growth. The UV channel will be used to quantify reduced nicotinamide adenine dinucleotide (NADH), a widely recognized indicator of cellular metabolic activity and intracellular redox state[66–68], which exhibits two characteristic absorption peaks at 240 and 360 nm[67,69]. The UV channel will also be employed to measure protein aggregation[70]. Hence, the UV channel serves a dual purpose by qualitatively monitoring microbial metabolic activity through NADH measurements and assessing protein aggregation in human γD-crystallin solutions. Ground-based functional validation of the spectroscopy module was performed using *S. oneidensis* MR1 and *U. maydis* FB1 populations, along with bovine serum albumin (BSA) and human γD-crystallin protein solutions. Transmitted light intensity from each biowell is converted into electrical signals by the corresponding photodiodes and digitized using an analog-to-digital converter (ADC; ADS1115 by Texas Instruments) interfaced with the Raspberry Pi through the I²C communication protocol. Measurements were stored as timestamped CSV files.

The electrochemical module measures microbial metabolic activity by monitoring of EET using a three-electrode electrochemical set-up (Figure 2B). Only a few spaceflight studies have demonstrated the feasibility of employing electroactive microorganisms for bioelectrochemical applications under microgravity conditions. In 2019, the Micro-12 flight experiment conducted aboard the ISS investigated the effects of microgravity on the physiology of the model EET organism, *S. oneidensis* MR-1[71]. The study demonstrated enhanced insoluble ferrihydrite reduction in spaceflight compared to ground conditions. More recently, microbial fuel cell (MFC) experiments aboard the ISS demonstrated increased voltage and power generation by *S. oneidensis* biofilms under microgravity conditions[72]. While previous ISS microbial fuel cell experiments successfully demonstrated that *S. oneidensis* biofilms remain electrochemically active and capable of power generation under microgravity conditions, these systems primarily measured bulk electrical output and lacked the ability to precisely control electrode potentials or interrogate mechanistic EET processes. The MAEx electrochemical module expands upon these earlier studies through the implementation of a three-electrode potentiostat architecture capable of applying defined electrode potentials at the working electrodes to mimic mineral reduction environments encountered by electroactive microorganisms[73–76]. A custom four-channel potentiostat is designed to enable parallel electrochemical measurements across four independent electrochemical biowells integrated within the biochamber assembly. The MAEx potentiostat architecture is a modified version of previous Arduino-based JUAMI potentiostat design[77] and redesigned around a Raspberry Pi-based control system to enable autonomous operation and seamless integration with the overall payload electronics. The system incorporates differential amplifier circuitry together with a 12-bit DAC (MCP4921) and 16-bit ADC (ADS1115) for improved analog signal generation and acquisition, enabling programmable electrochemical techniques such as cyclic voltammetry and chronoamperometry. Integration of the in-house potentiostat within the central Raspberry Pi architecture enabled unified hardware and software control across the entire MAEx payload. Ground-based functional validation of the electrochemical module was performed by running cyclic voltammogram[78] using the standard $Fe(CN)_6^{3-}/Fe(CN)_6^{4-}$ redox couple. Further electrochemical validation was conducted by monitoring electrochemical growth of *S. oneidensis* MR-1 using a carbon-cloth based working electrode poised at +0.4 V vs SHE, a platinum (Pt) mesh as the counter electrode, and an Ag/AgCl (1 M KCl) reference electrode.

**3.3 Biochamber Design**

For ground-based validation, the biochamber platform consists of a custom-designed polycarbonate (PC) device. The wells are directly machined into the central PC substrate. An additional thin PC layer serves as the top cover to improve mechanical rigidity and structural stability during launch and microgravity operation (Figure 2B). The top PC cover will be secured to the biochamber PC body using socket-head screws and lock nuts, with a silicone rubber sheet placed in between the two structures to ensure a leak-proof assembly. Two biochamber configurations were developed for integration with different analytical modules of the payload. The first configuration (Supplementary Figure 1A) was optimized for the imaging and spectroscopy analytical modules and consists of a 3 × 3 array of culture wells, with each well accommodating approximately 2 mL of culture volume. The second configuration (Supplementary Figure 1B), referred to as the electrochemical biochamber, was designed for integration with the electrochemical analytical module and consists of a 2 × 2 array of culture wells, each accommodating approximately 3.4 mL of culture volume. The electrochemical biochamber incorporates a three-electrode electrochemical system consisting of a carbon cloth based working electrode (WE), a Pt mesh counter electrode (CE), and an Ag/AgCl (1M KCl) reference electrode (RE). This design was inspired by single-chamber three-electrode electrochemical reactor system that have successfully enriched subsurface electroactive microorganisms[73–75], and is suggested as a putative design for electrobiotechnological application for single species microbial culture[76]. In the present design, precision-machined openings in the PC top cover and silicone rubber layer enables direct electrode insertion into individual culture wells. For spaceflight, the electrode and PC top cover interface will be sealed using compression-assisted bonding to maintain fluid containment and mechanical stability under spaceflight conditions.

### 3.4 House-keeping Module

For future spaceflight operation of MAEx payload, the housekeeping module will be incorporated to monitor environmental parameters within the payload, including temperature and pressure (BMP280, Bosch Sensortec). Passive thermal control will be provided by multilayer insulation (MLI) while active thermal regulation may be achieved using a thermoelectric cooler (TEC; TEC1-12705, generic) to maintain the desired operating temperature (25-30°C). Our preliminary thermal simulation suggest that active thermal control may be necessary for required temperature control (See Section: Spaceflight Environmental Validation). During ground-based functional validation, the temperature was continuously monitored for experimental control, with discrete measurements recorded at the beginning of each experimental cycle and stored alongside corresponding experimental data. These environmental measurements will be integrated as metadata with the individual modules to ensure traceability of experimental conditions during long-term spaceflight operations.

### 3.5 Power Distribution System

The MAEx platform incorporates power regulation by using switching DC-DC buck converters, selected for their high efficiency and suitability for large input-to-output voltage conversion with minimal heat dissipation. The system accepts an input voltage in the range of 24-36 V and regulates it to stable output voltages of 12 V and 5 V. The 12 V rail supplies high-power components, including the potentiostat and LED arrays, while the 5 V rail powers the Raspberry Pi and associated electronics. Additional voltage requirements, such as the 25-30 V bias required for UV photodiodes, are provided through dedicated regulation circuits.

### 3.6 Control System and Data Handling

During spaceflight, the control logic will follow a cyclic execution scheme (Figure 3A), with each cycle defined by a fixed duration of approximately 30 minutes. At the start of each cycle, a global timer will be reset, and flag variables (F2 and F3) will be initialized to zero, indicating that the spectroscopy and electrochemical modules have not yet been executed within that iteration. The system will first acquire the housekeeping data, including temperature and pressure, which will be recorded at the beginning of each cycle. Following this, the imaging module will be executed, where LED illumination will be activated, an image will be captured, and illumination will be subsequently turned off.

The control sequence will then proceed to the spectroscopy module, which will operate through a nested loop. With this subroutine, photodiode measurements will be acquired repeatedly (n = 200) to improve measurement precision and enable estimation of mean and variance. A counter variable (C), initialized at the start of the subroutine, will be incremented after each iteration until the predefined number of acquisitions is reached. Upon completion of the spectroscopy measurements, the electrochemical module will be executed, during which experimental parameters are set, and current and voltage data are recorded using the in-house potentiostat.

Following the execution of each module, the system will check whether the 30-minute cycle duration has elapsed. If time remains within the cycle, the system proceeds to the next pending experiment based on flag status; if all experiments are completed, the system will remain idle until the cycle duration is reached. Upon completion of the cycle, the timer is reset and the sequence is repeated.

**Figure 3:** (A) Flowchart illustrating the planned sequential execution of experiments controlled by the Raspberry Pi (RPi), including the imaging module (Mod1), spectroscopic module (Mod2), and electrochemical module (Mod3). F2 and F3 represent flag variables indicating whether the spectroscopic and electrochemical experiments, respectively, were executed during a given iteration. Conditional timing blocks will enforce a 30 min experimental cycle using a universal timer that will be reset at the start of each acquisition cycle. (B) Flowchart illustrating the proposed independent temperature-control program maintaining payload temperature within the predefined operating range.

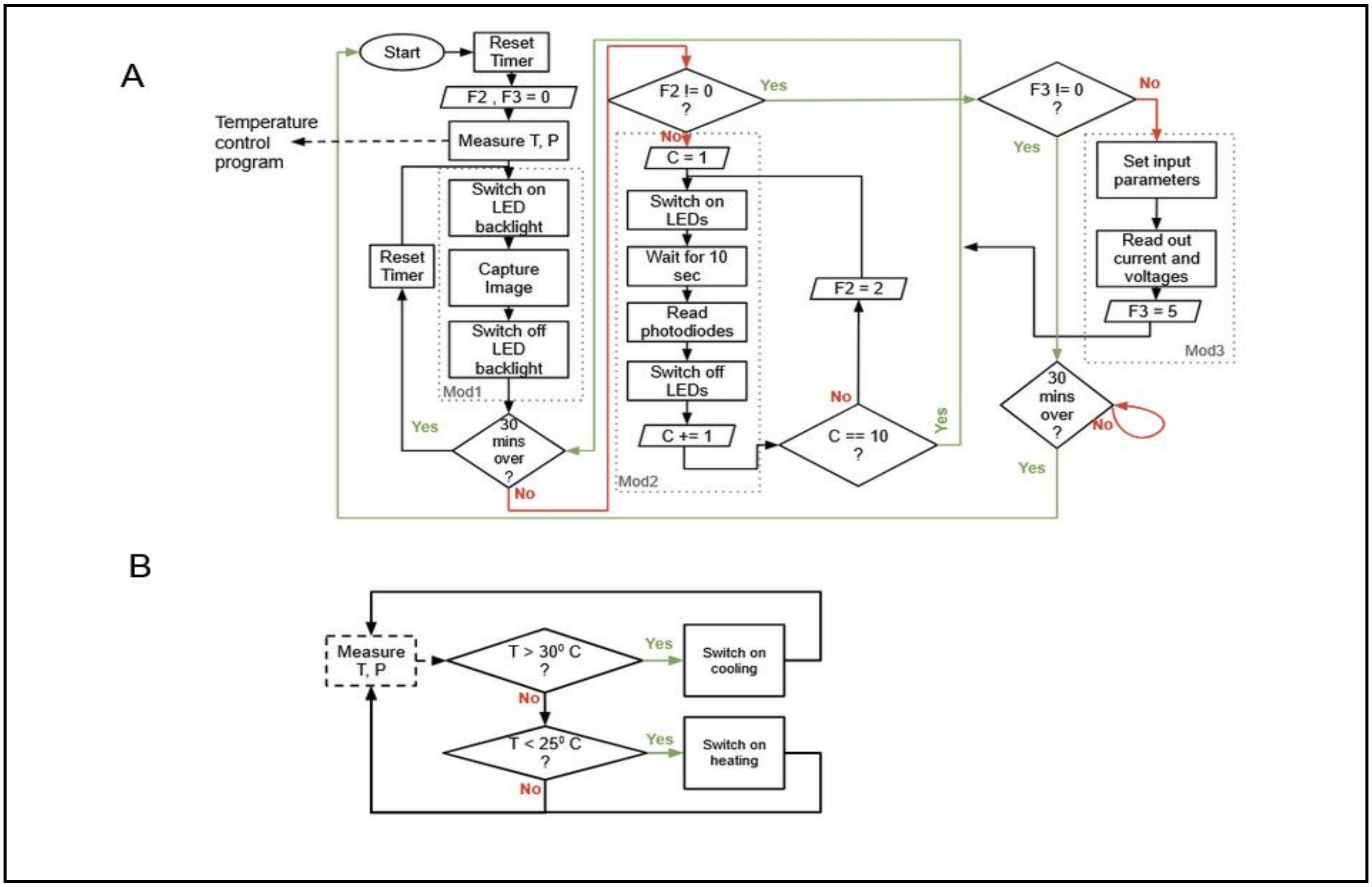

In parallel with the main control loop, an independent temperature regulation subroutine will operate continuously (Figure 3B). This routine will receive temperature input from the main program and will maintain thermal conditions within a predefined range (25-30°C) using a thermoelectric cooler (TEC). Heating or cooling will be activated when the temperature falls outside this range.

All data generated by the biological measurement modules and housekeeping sensors will be stored locally on the Raspberry Pi in structured formats (e.g., PNG for imaging data and CSV for sensor outputs). This data will be periodically transmitted to the parent satellite platform for eventual downlink to ground stations, ensuring reliable data retrieval and post-mission analysis.

### 3.7 Spaceflight Environmental Validation

For future operation in low Earth orbit (LEO), the MAEx payload must withstand the thermal, pressure, and operational constraints associated with spaceflight. These include extreme temperatures, ranging from -160°C during eclipse conditions to over +120°C under direct solar illumination, and near-vacuum pressures of ~$10^{-3}$–$10^{-5}$ hPa. The following analyses evaluate the preliminary thermal and structural responses of the proposed MAEx payload under environmental conditions representative of ISRO's POEM-4 platform.

For thermal control, we primarily hope to incorporate only a custom design multilayer insulation (MLI). Preliminary thermal simulations were conducted to estimate the payload's expected temperature range during orbital operation (see Methods). The model accounted for three primary external heat sources: direct solar radiation, Earth-reflected solar radiation (albedo), and infrared radiation emitted by the Earth. During sunlit orbital phases, all three heat sources contribute to the thermal balance, whereas during eclipse phases terrestrial infrared radiation becomes the dominant contributor. The simulations predicted an external platform temperature range of approximately -80°C to -10°C over a complete orbital cycle. When payload power dissipation was incorporated, the predicted temperature ranges increased to 20-70°C under idle conditions (~6 W) and 70-100°C under average operating conditions (~12 W). These temperatures exceed the desired operating range of 25-30°C, highlighting the need for additional thermal management measures. Hence, further thermal modelling is underway to evaluate passive thermal management strategies incorporating MLI (effective emissivity = 0.05; solar absorptivity = 0.01) and a dedicated radiator (infrared emissivity = 0.85; solar absorptivity = 0.01). Transient thermal simulations over repeated orbital cycles are underway to assess their effectiveness in maintaining the payload temperature within the desired operating range of 25-30°C. However, further discussions have led us to conclude that active temperature regulation will be vital to maintain the optimal temperature range.

**Figure 4:** Finite element modal simulations were performed to evaluate the dynamic structural response of the MAEx payload under launch-relevant conditions. The first six vibration modes and their corresponding natural frequencies are shown. The dominant structural mode (Mode 1) was identified at 218.72 Hz based on the highest effective mass and participation factor values. Since this frequency is substantially higher than the expected natural frequency range of the host spacecraft (70-90 Hz), the MAEx payload is expected to remain outside the resonance regime during nominal launch conditions. Forced vibrational analysis is still pending.

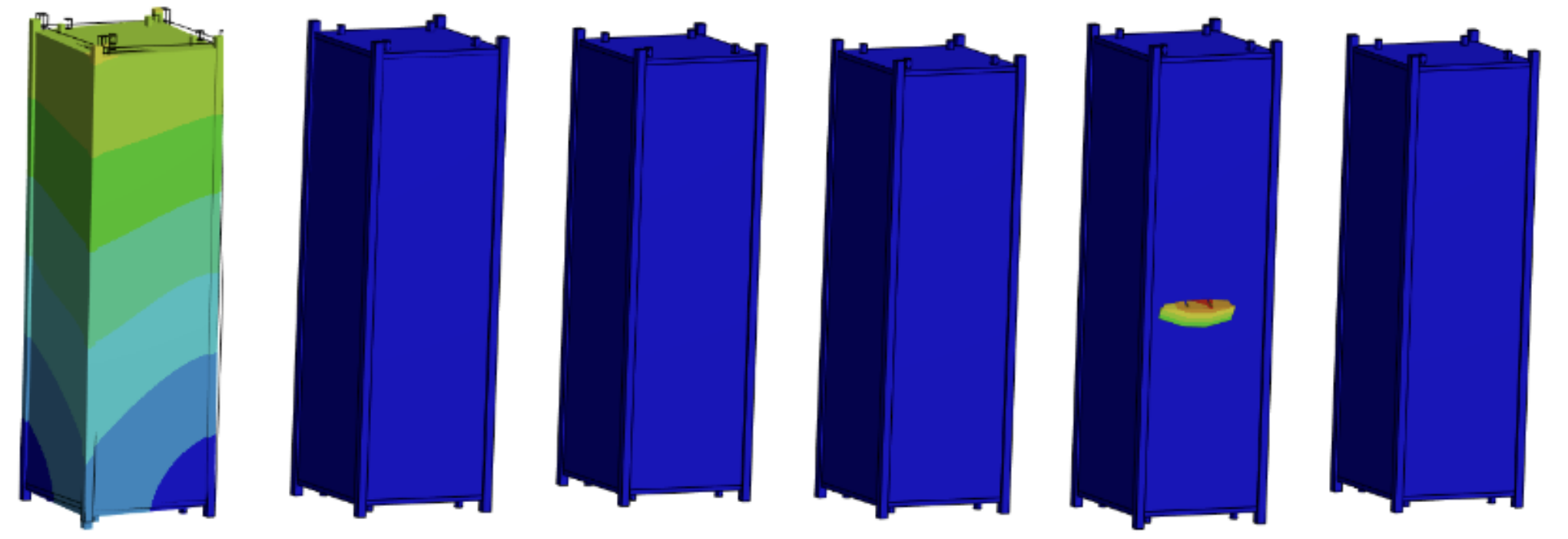

To evaluate the structural integrity of the MAEx payload under launch-induced vibrations, modal and random vibration analyses were performed using finite element simulations. Modal analysis identified the first and most dominant natural frequency of the structure at 218.72 Hz (Figure 4), which is higher than the estimated 70-90 Hz natural-frequency range of the host spacecraft. The dominant mode was identified based on its effective mass and modal participation factor. Random vibration simulations further revealed that the payload response was dominated by lateral bending modes at 467.8 Hz (X-axis) and 472.5 Hz (Y-axis), and an axial mode at 824.5 Hz (Z-axis), all of which fall within the high-energy region of the applied qualification spectrum (Supplementary Figure 2). The corresponding RMS acceleration responses were 40.5, 41.0, and 91.4 g along the X, Y, and Z axes, respectively, representing amplification factors of approximately 3.0, 3.0, and 6.8 relative to the input acceleration. The maximum predicted 3σ deformations were 0.86 mm and 0.87 mm in the X and Y directions, respectively, and 0.105 mm in the Z direction. These results indicate considerable dynamic amplification, particularly along the axial direction, and identify the structural modes that dominate the payload's response to launch-induced vibration. Although the first dominant MAEx natural frequency is separated from the estimated natural-frequency range of the host spacecraft, further assessment of structural stresses, mounting interfaces, and component-level acceleration limits will be necessary. Moreover, the qualification spectrum and mounting boundary conditions need to be verified against the applicable POEM-4/PSLV-C60 launch requirements, which were not readily available for this assessment.

To mitigate radiation-induced degradation during potential LEO operations, the MAEx payload will incorporate passive shielding and electronic fault-tolerance measures. The proposed 3-mm-thick Al 6061-T6 enclosure will provide baseline radiation shielding for the onboard electronics. However, the effectiveness of this shielding will depend on the orbital radiation environment, mission duration, and component-level shielding geometry. Although previous studies have demonstrated the potential radiation tolerance of commercial Raspberry Pi platforms[79–81], including the Raspberry Pi Zero W under Co-60 irradiation[82], these findings cannot be directly extrapolated to the Raspberry Pi 3 B+ and custom-designed PCBs employed in MAEx. Additional radiation-hardness assessment will be required to evaluate their susceptibility to total ionizing dose (TID) and single-event effects (SEE). These measures will help reduce the risk of data loss and operational interruptions during spaceflight.

Notwithstanding these qualification requirements for independent LEO operation, the current ground-validated MAEx payload architecture may be suitable for future integration into an existing crewed orbital laboratory, such as the International Space Station (ISS) and Tiangong Space Station, or planned orbital laboratory, such as Bharatiya Antariksh Station (BAS), where environmental control and infrastructure support facilitate biological experimentation.

## 4. Functional Validation of MAEx Analytical Modules

Demonstrating the functionality of the individual analytical modules under controlled laboratory conditions is a prerequisite for their subsequent integration and qualification for spaceflight. We therefore validated the imaging, spectroscopic, and electrochemical modules using three biological systems: *S. oneidensis* MR-1 and *U. maydis* FB1 to monitor growth and metabolic activity, and human γD-crystallin to assess protein aggregation.

### 4.1 Imaging Module

The imaging performance of the Raspberry Pi Camera Module 3 Wide was characterized under laboratory condition. The camera's field of view and image clarity was assessed at working distances ranging from 30 to 150 mm, in increments of 10 mm. The camera was positioned perpendicular to a calibrated grid, and images were acquired at each distance. Although different lens adjustment settings were initially investigated, the final imaging workflow employed the camera's autofocus mode, as fixed manual lens positions resulted in reduced image quality under low-light conditions arising during microbial growth. Based on the field-of-view characterization, a working distance range of 60-100 mm was identified (Supplementary Figure 3), providing a suitable balance between image coverage and focus stability.

Functional verification was performed by monitoring *S. oneidensis* MR-1 and *U. maydis* FB1 grown in Yeast Extract Peptone Dextrose (YPD) medium at 25°C. Two biochambers (with 3 x 3 biowells) were incubated in parallel: one was monitored using the imaging module, with the camera positioned 70 mm above the chamber, while the other was used for independent spectrophotometric measurements (SpectraMax M3 Multi-Mode Microplate Reader). No detectable growth was measured in the abiotic control (media only) during the 52-h incubation. Time-resolved imaging captured a reduction in light transmission due to progressive increase in culture turbidity (Supplementary Figure 4). The normalized mean intensity was calibrated against corresponding spectroscopic measurements acquired at 630 nm (reported as $OD_{600}$) at discrete time points (Supplementary Figure 5 and 6). Linear calibration yielded strong agreement between the negative logarithm of normalized mean image intensity and the corresponding $OD_{600}$ values for both organisms, with coefficients of determination ($R^2$) of 0.965 for *S. oneidensis* MR1 and 0.984 for *U. maydis* FB1 (Figure 5). The agreement between the imaging module and spectroscopic measurements, supports the use of the imaging module for both visual monitoring and quantitative assessment of microbial growth. A fisheye distortion effect was also observed in the acquired images; consequently, two Raspberry Pi cameras will be incorporated into the final payload configuration to improve image coverage and reduce optical distortion.

**Figure 5. Comparison of spectrophotometric and image-derived optical density measurements during microbial growth.** Growth of *S. oneidensis* MR1 and *U. maydis* FB1 observed using image-intensity-based analysis (dashed lines) and calibrated against spectroscopic readings at 630 nm (reported as $OD_{600}$; solid lines with markers). Shaded regions and error bars represent variability across replicate wells (±1 SD). The close agreement between the negative logarithm of normalized mean image intensity and spectroscopic measurements demonstrates the ability of the imaging module to quantitatively track microbial growth.

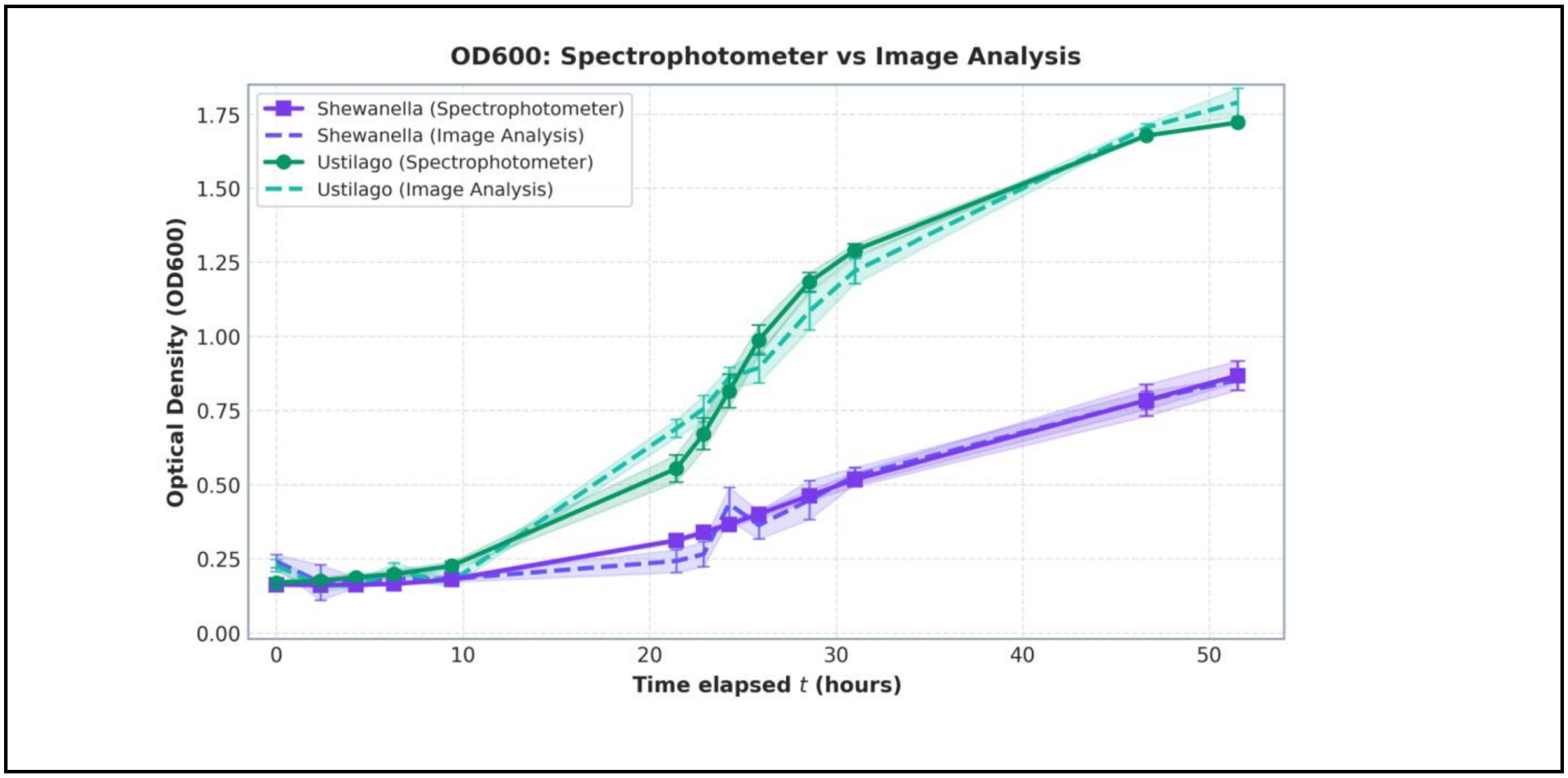

### 4.2 Spectroscopic Module

To enable absorbance measurements, the spectroscopy module employs red (630 nm) and ultraviolet (365 nm) LEDs together with wavelength-specific photodiodes; henceforth referred as red and UV channels. Prior to payload integration, the optical components were characterized to evaluate spectral performance, output stability, and temperature sensitivity under laboratory conditions (Supplementary Data). The measured spectra closely matched manufacturer specifications and confirmed suitability for spectroscopic measurements (Supplementary Figure 7).

Functional verification of the spectroscopy module was conducted using two experimental approaches: (1) endpoint measurements of late-exponential-phase cultures of *S.oneidensis* MR-1 and *U. maydis* FB1, alongside an aggregated human eye-lens γD-crystallin protein (2 mg/mL); and (2) time-resolved measurements to track microbial growth and protein aggregation. For the endpoint measurements, photodiode outputs were recorded using both red and UV channels for triplicate biowells for each biological sample, with 200 successive readings per biowell. In all the biological samples, we observed substantial decrease in normalized photodiode voltage of biological sample compared to abiotic control.

**Figure 6. Endpoint photodiode responses of the red and UV channel for biological system.** Photodiode output voltages were measured at 630 nm (red channel, red markers) and 365 nm (UV channel, blue markers) for the abiotic controls (triplicates) and corresponding biological samples (triplicates) of *S. oneidensis* MR-1, *U. maydis* FB1, and aggregated human γD-crystallin protein (2 mg/mL). Measurements were performed in triplicates, with 200 successive readings acquired per replicate. Error bars represent variability across replicates.

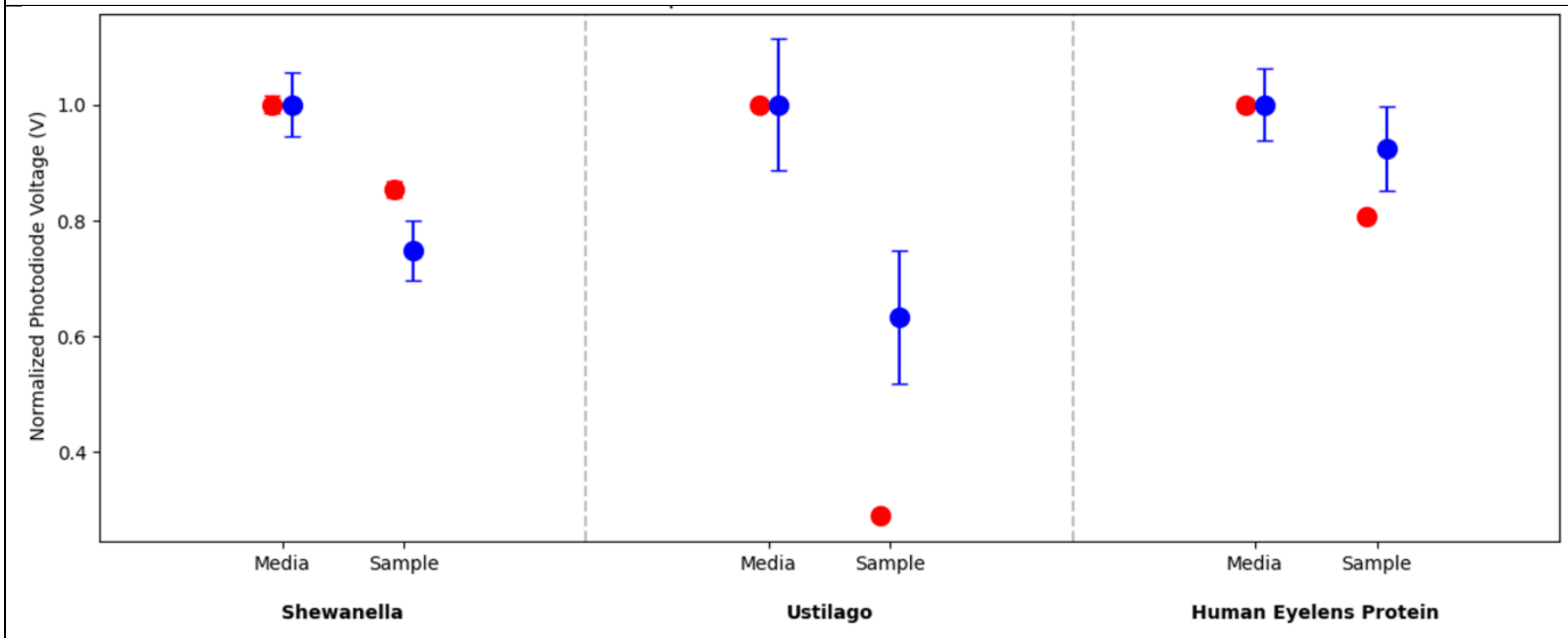

Additionally, time-resolved measurements (n=3) to monitor changes in optical transmission during microbial growth and protein aggregation were carried out. For all model biological systems (*S. oneidensis* MR-1 inoculated in LB agar, *U. maydis* FB1 cultivated in YPD medium and human γD-crystallin protein in phosphate buffer) the red-channel photodiode measurements showed progressive decrease (corresponding to increased turbidity of the sample) across the 2-day growth/aggregation period. Significant progressive decrease was also seen for the UV-channel photodiode measurements for *S. oneidensis* MR1, reflecting UV attenuation associated with increased NADH concentration, but the decrease was less pronounced for *U. maydis* FB1 and human γD-crystallin. The limited UV response in *Ustilago* is likely due to the strong UV attenuation by the YPD medium, resulting in a low output voltage of the UV photodiode and decreased signal-to-noise ratio. In γD-crystallin the limited UV response is likely due to the low concentration (2 mg/ml) of the γD-crystallin used for the experiment (for comparison, BSA protein showed significant UV response above 10 mg/ml;

see Supplementary Figure 11). Nevertheless, these results demonstrate the ability of the spectroscopy module to capture time-dependent changes in optical transmission during microbial cultivation (or protein aggregation) using complementary red and UV channels.

**Figure 7. Normalized photodiode voltage responses of biological samples under continuous UV and red LED illumination over a period of 2 days.** Time-course measurements are shown for *S. oneidensis* MR1 (left), *U. maydis* FB1 (center), and human γD-crystallin protein (right), exposed to UV LED illumination (top row, blue) and red LED illumination (bottom row, red). For each sample (triplicate), 200 raw data points were recorded during each measurement iteration and subsequently averaged over 5-minute intervals. Lightly shaded points represent all individual raw measurements, while solid lines indicate the mean responses calculated from the averaged data points. The corresponding shaded regions around the solid lines represent the standard error of the mean.

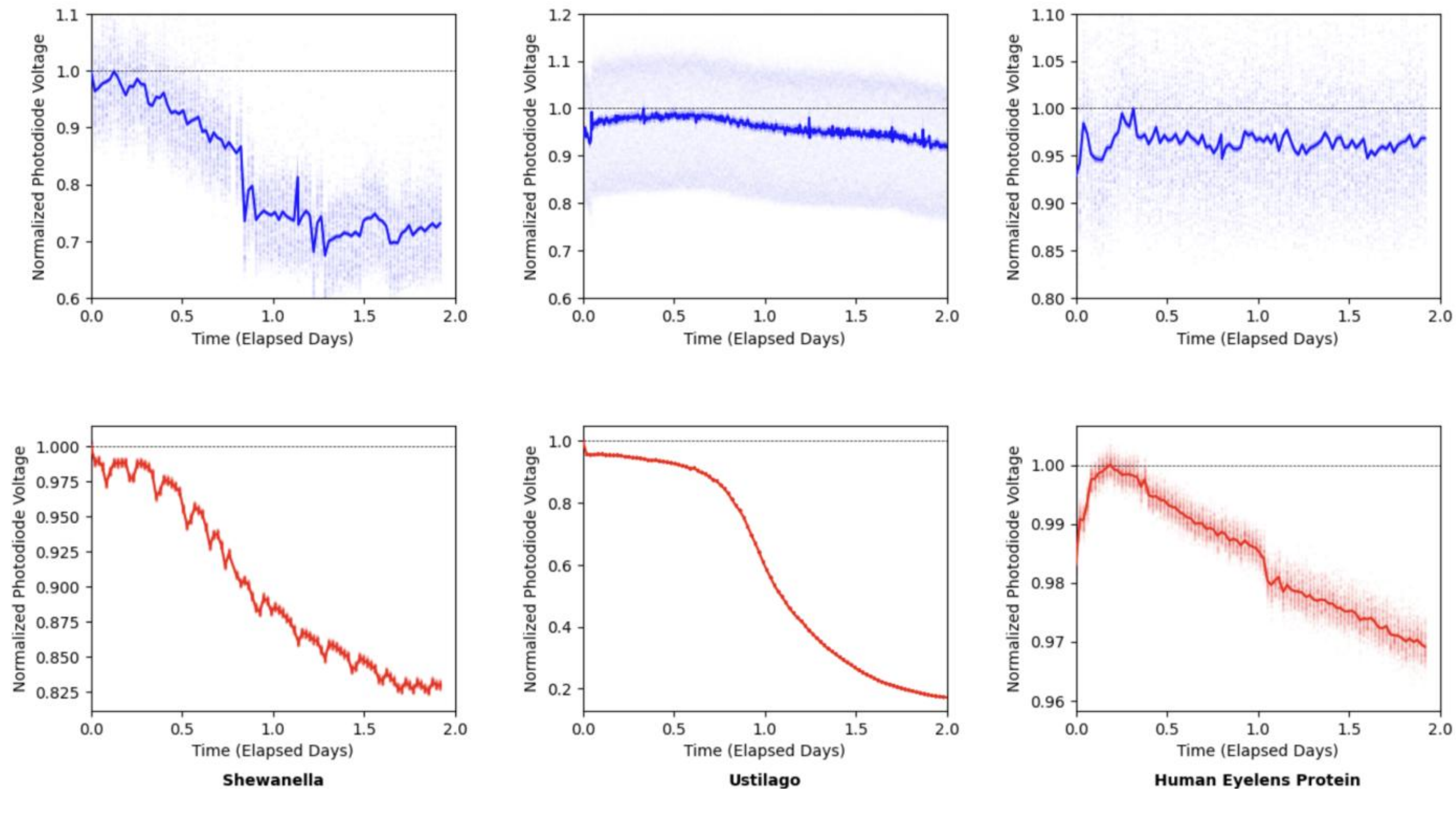

## 4.3 Electrochemical Module

The performance of the in-house PCB potentiostat was evaluated by checking long term stability of the working potential at +0.4 V vs Ag/AgCl (1M KCl) in a 3-electrode electrochemical system. We observed that the potential varied by 0.1% whereas the measured current fluctuated by <20 µA over a 60-hour measurement period. Next, we performed cyclic voltammetry (CV) measurements with a standard ferri/ferrocyanide redox system using a 3-electrode electrochemical system using 5 mM potassium ferricyanide in 0.1 M $KNO_3$ electrolyte (Figure 10). The recorded cyclic voltammograms exhibited characteristic redox responses associated with the $Fe(CN)_6^{3-}/Fe(CN)_6^{4-}$ couple with a midpoint potential ($E_{1/2}$) of ~0.265 V vs. Ag/AgCl. This value is consistent with previously reported midpoint potentials, usually varying due to electrolyte composition, ionic strength, electrode material, and uncompensated resistance effects[78,83,84].

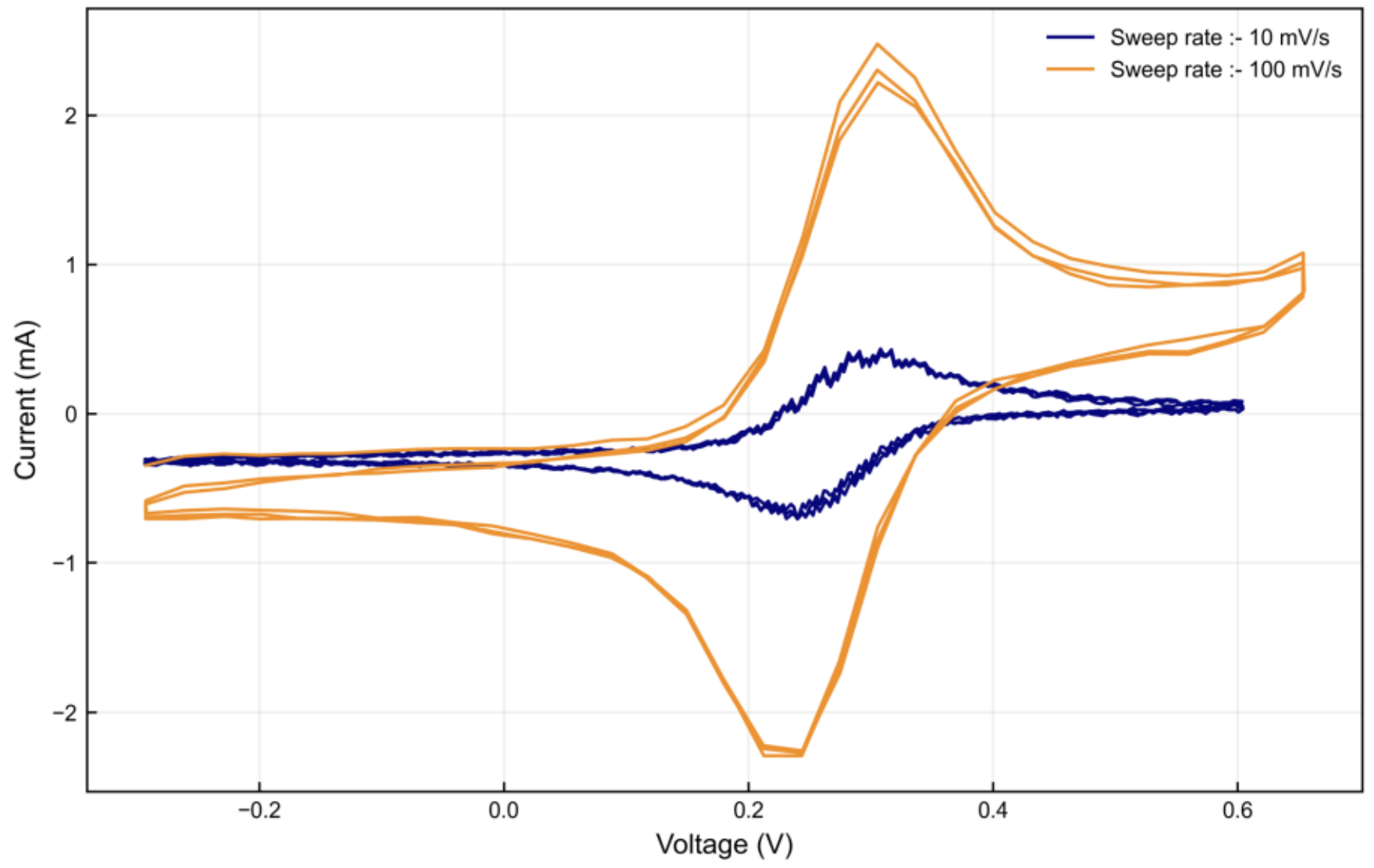

**Figure 8:** Electrochemical characterization of the MAEx potentiostat module. Cyclic voltammetry measurements (cycles n =3) were performed using 5 mM $K_3Fe(CN)_6$ dissolved in 0.1 M $KNO_3$ electrolyte to validate the electrochemical performance of the custom in-house poteniostat.

The electrochemical measurements showed a consistent increase in anodic current density during *S.oneidensis* MR-1 cultivation across two independent experiments performed using the custom in-house potentiostat. During microbial electrochemical incubation, we observed a larger signal variability with the in-house custom potentiostat relative to the commercial instrument. One possible contributing factor is the difference in data acquisition: the custom potentiostat averages 10 readings at each measurement point, whereas the internal averaging settings of the MedPStat 2.0 were not accessible for direct comparison. In-house custom potentiostat exhibited lower signal variability during initial validation using the $[Fe(CN)_6]^{3-}/[Fe(CN)_6]^{4-}$ redox couple (Figure 10) compared to microbial electrochemical incubation. This difference is due to the distinct nature of the two electrochemical systems. Ferricyanide undergoes a well-defined solution-phase redox reaction involving direct electron transfer at the electrode, whereas microbial current generation involves more complex set of biological and interfacial processes[85,86]. Thus, the greater variability observed during microbial cultivation may reflect a combination of biological variability and data-acquisition settings.

The maximum current densities reached approximately 1.3 A/m² and 1.0 A/m² in the two custom potentiostat runs, respectively, compared with approximately 0.2 A/m² measured using the commercial potentiostat. Previously reported, current density measurements for *S.oneidensis* MR1 ranges from 0.1- 3.0 A/m$^2$ depending on its environment[87,41]. These variation reflect difference in metabolic activity, outer-membrane cytochrome expression, cell-electrode interactions, and media composition[88–91]. Given that the biowells were supplied with complex Luria-Bertani (LB) medium, a relatively high anodic current densities are expected, as observed for in-house potentiostats. In addition to observing anodic current with constant potential applied, we also performed slow scan CV to assess presence of any redox reactions and whether the biofilm was under turnover conditions (actively oxidizing a carbon source). We observed a catalytic oxidation at midpoint potential of ~ 0.05 V vs. SHE (Supplementary Figure. 12), similar to a recent observation of *S. oneidensis* MR1 biofilm[92]. However, it is important to note previous study also report that the electrode-attached *S. oneidensis* MR1 biofilm exhibits catalytic peaks at -0.11 V, -0.24 V, and +0.21 V vs SHE associated with free-flavin, outer-membrane cytochrome bound flavins, and outer membrane cytochromes, respectively[89]. Considering these observations, the redox peak detected in our study may reflect the combined contributions of multiple redox processes associated with the oxidation of rich growth medium by the electrode-attached biofilm.

Taken together, the cyclic voltammetry (CV) measurements using the potassium $[Fe(CN)_6]^{3-}/[Fe(CN)_6]^{4-}$ redox couple and the chronoamperometric (CA) and cyclic voltammetry (CV) measurements of *S. oneidensis* MR-1 electrocultivation demonstrate the functional performance of the MAEx electrochemical module.

**Figure 9.** Comparison of anodic current density measured during Shewanella oneidensis MR-1 cultivation using the custom PCB potentiostat and commercial MTX MedPStat 2.0. Current density was monitored over 60-h using a three-electrode electrochemical cell. The two custom potentiostat measurements represent independent experimental replicates. Raw current data (light shades) and smoothed lines (dark shades) are shown for the custom potentiostat measurements (yellow and brown solid lines), alongside the commercial potentiostat (MedPStat 2.0; blue solid line). Control experiments showed no appreciable current generation (not included).

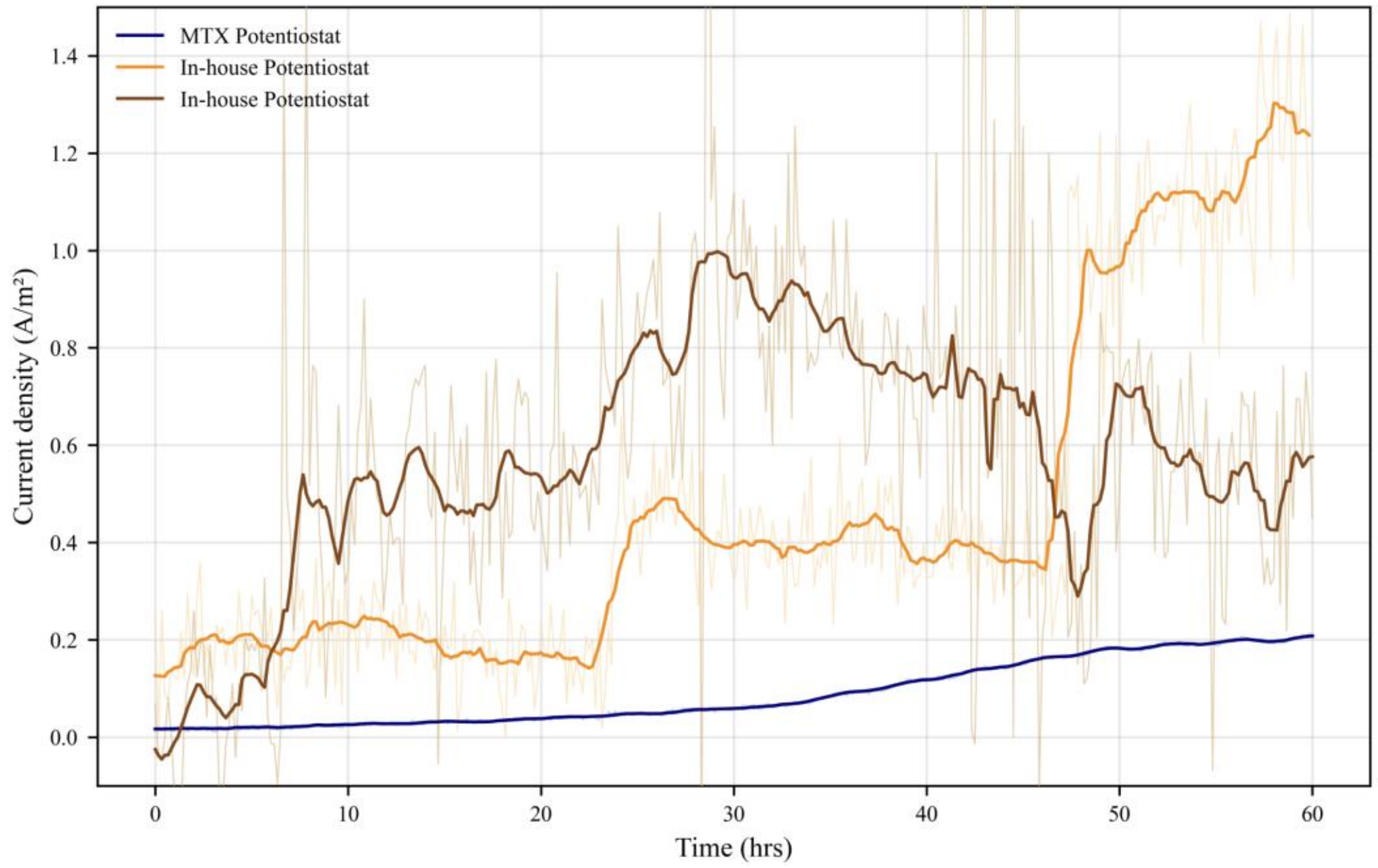

## 6. Conclusion

In this study, we present the design and ground-based functional validation of the Modular Astrobiological Experiments (MAEx) payload, a 3U bioanalytical platform for monitoring of biological systems under spaceflight conditions. MAEx integrates imaging, UV-Vis spectroscopy, and electrochemical sensing within a small-satellite-compatible architecture. Ground-based experiments demonstrated the functionality of the individual analytical modules, including camera-based visualization and quantitative monitoring of microbial growth, LED-photodiode UV-Vis spectroscopy based qualitative investigation of microbial growth and protein-aggregation, and microbial electrochemistry based qualitative investigation of microbial respiration. This work establishes the ground-based functional validation of the proposed MAEx payload platform. A complete assessment of unresolved challenges, including long-term autonomous operation, sensor drift, biofouling, contamination control, radiation susceptibility, fluid handling under microgravity, and payload qualification, required to make MAEx spaceflight ready, is currently underway. The modular design of MAEx provides an opportunity to incorporate into current or future crewed orbital research facilities. Its ground-validated architecture offers opportunities for biological investigations aboard operational platforms such as the ISS and Tiangong Space Station, as well as emerging facilities such as the Bharatiya Antariksh Station (BAS), subject to the relevant integration and safety requirements.

Earlier biological CubeSat missions, including GeneSat-1, PharmaSat, O/OREOS, EcAMSat, and BioSentinel, primarily relied on optical measurements to monitor microbial growth, metabolism, and biological responses. In parallel, the development of a multimodal electrochemical-optical CubeSat platform, incorporating pH and sodium ion-selective sensors alongside optical absorbance measurements, demonstrated the feasibility of combining complementary sensing techniques within a compact payload. Building on these approaches, MAEx integrates imaging, spectroscopy, and electrochemical interrogation within a single payload architecture. This combination offers the potential to correlate observable changes in cellular growth and optical properties with underlying electrochemical activity. Inclusion of electrochemical tools provides an additional means of investigating microbial electron-transfer processes that are not readily resolved through optical measurements alone. Thus, MAEx provides a complementary approach to existing biological satellite platforms, expanding the range of biological processes that can be investigated using compact, integrated payloads.

Beyond its immediate application in astrobiology and space biology, the multiple analytical tools incorporated in architecture of MAEx has potential relevance to the development of biological manufacturing and bioprocessing capabilities in space. Future space-based biomanufacturing systems will employ mutiple microorganisms or biological materials for the production of biomolecules, biopolymers, biomaterials, and other high-value products. Monitoring such processes requires complementary investigations on biomass accumulation, biochemical changes, and metabolic or electrochemical activity. By combining different analytical techniques, MAEx provides a potential framework for monitoring these parameters within a compact experimental platform. Importantly, the MAEx payload architecture and its analytical capabilities are independent of the specific biological model systems used in this study. The selected organisms and biomolecules serve as representative test systems for validating the functional operation of the individual payload modules. The platform can accommodate other biological systems relevant to future space biology and biomanufacturing investigations. Thus, the design and ground-validation results provide an engineering foundation for future payloads investigating biological production, microbial resource utilization, and bioregenerative life-support processes during long-duration operations in low Earth orbit or deep-space exploration.

## 7. Methods

### Microbial Strains and Culture Conditions

The bacterial strain *Shewanella oneidensis* MR-1 was a generous donation from Mohamed El-Naggar, University of Southern California, USA. The fungal strain *Ustilago maydis* FB1 was kindly provided by Saravanan Matheswaran, Indian Institute of Technology Kanpur, India. Traditionally, *S. oneidensis* is cultured in Luria Broth (LB) medium, whereas *U. maydis* is routinely grown in Yeast Extract Peptone Dextrose (YPD) medium. YPD medium was prepared using yeast extract (10 g/L), peptone (20 g/L), and dextrose (20 g/L) dissolved in 1L Milli-Q water, followed by sterilization by autoclaving at 121 °C for 20 min. Meanwhile, Luria broth (LB) medium was prepared using tryptone (10 g/L), yeast extract (5 g/L), and sodium chloride (10 g/L) dissolved in 1 L of Milli-Q water, followed by sterilization by autoclaving at 121 °C for 20 min. Starter cultures were initiated from glycerol stocks stored at -80$^{o}$C and grown overnight under aerobic conditions. Both cultures were incubated statically at 30°C without shaking. Overnight cultures were subsequently diluted into respective media to an initial optical density ($OD_{600}$) of approximately 0.1 prior to loading into MAEx biochambers.

### Preparation of Protein Aggregates

For initial UV-Vis spectroscopy module validation, we prepared a 40 mg/mL Bovine Serum Albumin (BSA) stock solution by dissolving 400 mg of BSA (Sigma-Aldrich) in 10 mL of phosphate buffer (pH 7.4). Mix gently until fully dissolved. The stock solution was subsequently diluted with the same buffer to obtain final BSA concentrations of 0.1, 1, 4, 8, 10, and 20 mg/mL. BSA solutions were used to assess the minimum concentration at which the UV channel could reliably detect changes associated with protein aggregation. A minimum detectable concentration of 4 mg/mL was identified when a sufficient number of measurements (n = 200) were collected.

For *in vitro* aggregation experiments, we were able to purify human γD crystallin (HγD-Crys) to a final concentration of 2 mg/mL in phosphate buffer maintained at low pH conditions (pH 3.5). The acidic environment was used to induce gradual protein aggregation under controlled conditions. Samples were incubated statically at 25°C and monitored over a period of up to 35 days.

### Imaging module analysis

Aliquots of diluted cultures were aseptically transferred into sterile biochambers integrated within the MAEx payload. Each chamber contained 0.5 mL of YPD agar supplemented with 100 µL of microbial culture and was sealed to minimize contamination. Control chambers containing sterile YPD agar with 100 µL of Milli-Q water were included for baseline signal correction. Images of the payload biochambers were acquired at 10 min intervals throughout the experiment.

Images acquired during the incubation period were analyzed using OpenCV (cv2) to monitor changes in microbial growth based on transmitted light intensity. Plate movement and rotation were tracked frame-by-frame using reference marks to maintain consistent well positioning. For each well, a circular region of interest (ROI), corresponding to 90% of the detected well radius, was selected to minimize interference from reflections, shadows, and meniscus effects near the well edges (Supplementary Figure 3). The mean pixel intensity within this ROI was calculated to quantify transmitted light, as previously reported[93,94]. To account for initial differences in illumination and well positioning, sample intensities were first adjusted to match the corresponding control intensity at $t = 0$. The adjusted intensities were subsequently normalized against the control at each time point to compensate for temporal variations in illumination (Supplementary Figure 4). The normalized intensity was converted to an image-derived optical density ($OD_{image}$) using the negative base-10 logarithm. Finally, these values were calibrated against laboratory spectrophotometer measurements at 630 nm (reported as $OD_{600}$; Supplementary Figure 5) using linear regression to obtain calibrated optical density estimates.

### Spectroscopy module analysis

The emission spectra of the red, white, and UV LEDs and normalized spectral response curves of the corresponding photodiodes were characterized (Supplementary Figure 6). Details of temperature and signal stability is provided as Supplementary Data (Supplementary Figure 7-9),

For time-series measurements, microbial growth experiments were initiated by loading each bio-well with 0.5 mL of agar-based medium, followed by inoculation with 100 µL of diluted microbial culture. Wells containing sterile Milli-Q water served as negative controls. For protein aggregation experiments, 0.5 mL of purified human γD-crystallin (HγD-Crys) solution was loaded into each bio-well, while control wells contained HγD-Crys in phosphate buffer (pH 7.0) under non-aggregating conditions. The biochambers were maintained at 25 °C throughout the experiments.

For single-endpoint measurements, *S. oneidensis* MR-1 and *U. maydis* FB1 were grown in the respective biochambers, and measurements were acquired between 24 and 36 h of incubation, following a visual confirmation of growth. Similarly, aggregated HγD-Crys samples were prepared in triplicate bio-wells and loaded into the spectroscopy module. Separate biochambers containing the corresponding sterile media (LB and YPD) or phosphate buffer served as abiotic controls.

UV–Vis measurements were acquired using the red and UV channels. For time-series experiments, measurements were collected at 30-min intervals, with 200 readings acquired per channel at each time point. For single-endpoint experiments, measurements were acquired once after the respective incubation period, with 200 readings collected per channel per well.

**Electrochemical module analysis**

Cyclic voltammetry (CV) measurements were performed using 50 mL of an electrolyte solution containing 5 mM analyte and 0.1 M $KNO_3$ as the supporting electrolyte. The electrochemical measurements were carried out using an in-house potentiostat. Cyclic voltammograms were recorded at two scan rates, 10 mV/s and 100 mV/s, with three consecutive cycles at each scan rate to investigate the electrochemical response at different potential sweep rates. The data acquisition was configured with a sample interval of 300 ms and a data averaging factor of 6.

For electrochemical cultivation of *S. oneidensis* MR1, four independent electrochemical biowells containing 2 mL LB agar (1.5%) were prepared for comparative electrochemical analysis. Three of these biowells were supplemented with 1 mL of an actively growing *S. oneidensis* MR1 culture. In contrast, the abiotic negative control consisted of 1 mL sterile LB media (autoclaving at 121°C for 20 min). Each biowell was configured as a three-electrode electrochemical cell. Two of the incubated biowells were connected to a in-house potentiostat and the third one was connected to commercial single channel potentiostat (MedPStat 2.0, MTXLabs, India). The fourth biowell (acted as abiotic control). The electrochemical setup consisted of a carbon cloth working electrode, an Ag/AgCl (1 M KCl) reference electrode, and a platinum wire counter electrode. The exposed surface area of the working electrode was 0.5 $cm^2$. Chronoamperometric measurements were performed by poising the working electrode at +0.2 V versus Ag/AgCl throughout the experiment. Anodic current was recorded at 10-min intervals over a 60-h incubation period, with each measurement averaged over 10 readings.

**Thermal Simulation**

The STK-SEET thermal model estimates the mean steady-state temperature of a spacecraft using simplified thermal balance equations while approximating the vehicle as a single isothermal node with user-defined bulk thermal properties. The simulations incorporated three primary external heat sources: direct solar radiation, Earth-reflected solar radiation (albedo), and terrestrial infrared radiation emitted by the Earth. During sunlit orbital phases, all three heat sources contribute to the thermal balance, whereas during eclipse conditions the terrestrial infrared component remains the dominant external heat source. Representative spacecraft thermal parameters used in the simulation included an Earth albedo value of 0.34, material emissivity of 0.924, material absorptivity of 0.14, and a vehicle cross-sectional area of 0.0185 $m^2$. Internal power dissipation was assumed to be 0, 6, and 12 W for the preliminary environmental estimation. The thermal model geometry was approximated using a plate-based representation aligned relative to the Sun vector within the STK-SEET framework. Further thermal modelling is underway to incorporate a multilayer insulation (MLI) with effective emissivity of 0.05; solar absorptivity of 0.05, with an infrared emissivity of 0.85 and solar absorptivity of 0.05.

**Vibrational Simulation**

Random vibration analysis was performed using ANSYS Mechanical (2023 R1) to evaluate the structural response of the MAEx payload under broadband launch-induced vibrations. A mode-superposition approach was used, with the payload constrained at its mounting standoffs. The qualification-level power spectral density (PSD) profile (Supplementray Table 2), corresponding to an overall acceleration of 13.5$g_{rms}$ and a test duration of 2 min per axis, was applied as base acceleration independently along the X, Y, and Z axes. An acceptance-level profile of 9.0 $g_{rms}$, with the same spectral shape and a duration of 1 min per axis, was also considered. The input PSD profiles were defined at discrete frequencies of 20, 60, 250, 1000, and 2000 Hz, with the corresponding acceleration spectral densities specified for each level. The resulting acceleration response PSDs, RMS accelerations, and maximum three-sigma (3σ) deformations were evaluated along each axis to identify the dominant structural modes and assess the payload's dynamic response (Supplementary Table 3).

## 8. References

1. Prasad, B. *et al.* How the space environment influences organisms: an astrobiological perspective and review. *International Journal of Astrobiology* **20**, 159–177 (2021).

2. Lu, Y., Shao, Q., Yue, H. & Yang, F. A Review of the Space Environment Effects on Spacecraft in Different Orbits. *IEEE Access* **7**, 93473–93488 (2019).

3. Thirsk, R., Kuipers, A., Mukai, C. & Williams, D. The space-flight environment: the International Space Station and beyond. *CMAJ* **180**, 1216–1220 (2009).

4. Averesch, N. J. H. *et al.* Microbial biomanufacturing for space-exploration—what to take and when to make. *Nat Commun* **14**, 2311 (2023).

5. McNulty, M. J. *et al.* Molecular pharming to support human life on the moon, mars, and beyond. *Critical Reviews in Biotechnology* **41**, 849–864 (2021).

6. Koehle, A. P., Brumwell, S. L., Seto, E. P., Lynch, A. M. & Urbaniak, C. Microbial applications for sustainable space exploration beyond low Earth orbit. *npj Microgravity* **9**, 47 (2023).

7. Santomartino, R. *et al.* Toward sustainable space exploration: a roadmap for harnessing the power of microorganisms. *Nat Commun* **14**, 1391 (2023).

8. Fountain, L. L. *et al.* Expanding frontiers: harnessing plant biology for space exploration and planetary sustainability. *New Phytologist* **249**, 656–669 (2026).

9. Maffei, M. E. *et al.* The physiology of plants in the context of space exploration. *Commun Biol* **7**, 1311 (2024).

10. Bunchek, J. M. *et al.* Pick-and-eat space crop production flight testing on the International Space Station. *Journal of Plant Interactions* **19**, 2292220 (2024).

11. Kordyum, E. & Hasenstein, K. H. Plant biology for space exploration – Building on the past, preparing for the future. *Life Sciences in Space Research* **29**, 1–7 (2021).

12. Marotta, D. *et al.* Biomanufacturing in low Earth orbit: A paradigm shift. *Stem Cell Reports* **20**, (2025).

13. Sharma, A. *et al.* Biomanufacturing in low Earth orbit for regenerative medicine. *Stem Cell Reports* **17**, 1–13 (2022).

14. Guan, D., Li, L., Zhu, Z. & Zheng, X. Made in space, used on Earth: Toward a new era of space biopharmaceuticals under microgravity. *Innovation* **0**, (2026).

15. Marshall Porterfield, D. *et al.* Critical investments in bioregenerative life support systems for bioastronautics and sustainable lunar exploration. *npj Microgravity* **11**, 57 (2025).

16. Fu, Y. *et al.* How to Establish a Bioregenerative Life Support System for Long-Term Crewed Missions to the Moon or Mars. *Astrobiology* **16**, 925–936 (2016).

17. Dempster, W. F. *et al.* Technical review of the Laboratory Biosphere closed ecological system facility. *Advances in Space Research* **34**, 1477–1482 (2004).

18. Poughon, L., Farges, B., Dussap, C. G., Godia, F. & Lasseur, C. Simulation of the MELiSSA closed loop system as a tool to define its integration strategy. *Advances in Space Research* **44**, 1392–1403 (2009).

19. Younger, A. K. Orbital biomanufacturing: the key to space resiliency, operational persistence, and ethical sustainability. *npj Microgravity* **12**, 35 (2026).

20. Cockell, C. S. *et al.* BioMoon: a concept for a mission to advance space life sciences and astrobiology on the Moon. *Discov Sp* **128**, 5 (2024).

21. Berliner, A. J. *et al.* Towards a Biomanufactory on Mars. *Front. Astron. Space Sci.* **8**, (2021).

22. Zhang, Y. *et al.* NASA's Ground-Based Microgravity Simulation FacilityMicrogravity simulation facility. in *Plant Gravitropism: Methods and Protocols* (ed. Blancaflor, E. B.) 281–299 (Springer US, New York, NY, 2022). doi:10.1007/978-1-0716-1677-2_18.

23. Brungs, S. *et al.* Facilities for Simulation of Microgravity in the ESA Ground-Based Facility Programme. *Microgravity Sci. Technol.* **28**, 191–203 (2016).

24. Simonsen, L. C., Slaba, T. C., Guida, P. & Rusek, A. NASA's first ground-based Galactic Cosmic Ray Simulator: Enabling a new era in space radiobiology research. *PLOS Biology* **18**, e3000669 (2020).

25. Norbury, J. W. *et al.* Galactic cosmic ray simulation at the NASA Space Radiation Laboratory. *Life Sciences in Space Research* **8**, 38–51 (2016).

26. La Tessa, C., Sivertz, M., Chiang, I.-H., Lowenstein, D. & Rusek, A. Overview of the NASA space radiation laboratory. *Life Sciences in Space Research* **11**, 18–23 (2016).

27. Ferranti, F., Del Bianco, M. & Pacelli, C. Advantages and Limitations of Current Microgravity Platforms for Space Biology Research. *Applied Sciences* **11**, 68 (2021).

28. Acres, J. M., Youngapelian, M. J. & Nadeau, J. The influence of spaceflight and simulated microgravity on bacterial motility and chemotaxis. *npj Microgravity* **7**, 1–11 (2021).

29. Kitts, C. *et al.* Flight Results from the GeneSat-1 Biological Microsatellite Mission. *The Proceedings of 21st Annual AIAA/USU Conference on Small Satellite* https://digitalcommons.usu.edu/smallsat/2007/all2007/69 (2007).

30. Ricco, A. J. *et al.* PharmaSat: drug dose response in microgravity from a free-flying integrated biofluidic/optical culture-and-analysis satellite. in *Microfluidics, BioMEMS, and Medical Microsystems IX* vol. 7929 217–225 (SPIE, 2011).

31. Ehrenfreund, P. *et al.* The O/OREOS mission—Astrobiology in low Earth orbit. *Acta Astronautica* **93**, 501–508 (2014).

32. Padgen, M. R. *et al.* *EcAMSat* spaceflight measurements of the role of σs in antibiotic resistance of stationary phase *Escherichia coli* in microgravity. *Life Sciences in Space Research* **24**, 18–24 (2020).

33. Massaro Tieze, S., Liddell, L. C., Santa Maria, S. R. & Bhattacharya, S. BioSentinel: A Biological CubeSat for Deep Space Exploration. *Astrobiology* **23**, 631–636 (2023).

34. Śniadek, P. *et al.* Autonomous, miniature research station (lab-payload) for the nanosatellite biological mission: LabSat. *Sci Rep* **15**, 30898 (2025).

35. Lingam, N. *et al.* NASA's BioSentinel deep space CubeSat mission: successes and lessons learned. *Acta Astronautica* **236**, 188–193 (2025).

36. Lambacher, A., Akiyama, K., Lingam, N., Shuman, D. & Santa Maria, S. R. LEIA: NASA's first biological mission on the lunar surface since 1972. *Acta Astronautica* **231**, 148–152 (2025).

37. Kim, S., Park, S. & Pak, J. J. Multi-Modal Multi-Array Electrochemical and Optical Sensor Suite for a Biological CubeSat Payload. *Sensors* **24**, (2024).

38. Griffa, M., Di Giovanni, A., Santaera, G., Stefanini, C. & Romano, D. Design and evaluation of a miniaturized bioregenerative microecosystem for CubeSat missions. *Acta Astronautica* **241**, 278–292 (2026).

39. Ikeda, S. *et al.* Shewanella oneidensis MR-1 as a bacterial platform for electro-biotechnology. *Essays in Biochemistry* **65**, 355–364 (2021).

40. Holloman, W. K., Schirawski, J. & Holliday, R. Towards understanding the extreme radiation resistance of Ustilago maydis. *Trends in Microbiology* **15**, 525–529 (2007).

41. Bretschger, O. *et al.* Current production and metal oxide reduction by Shewanella oneidensis MR-1 wild type and mutants. *Applied and Environmental Microbiology* **73**, 7003–7012 (2007).

42. Shi, L. *et al.* Extracellular electron transfer mechanisms between microorganisms and minerals. *Nature Reviews Microbiology* **14**, 651–662 (2016).

43. Kim, B. *et al.* Progress and Prospects for Applications of Extracellular Electron Transport Mechanism in Environmental Biotechnology. *ACS EST Eng.* **4**, 1520–1539 (2024).

44. Castelein, S. M. *et al.* Iron can be microbially extracted from Lunar and Martian regolith simulants and 3D printed into tough structural materials. *PLOS ONE* **16**, e0249962 (2021).

45. Cortesão, M., Schütze, T., Marx, R., Moeller, R. & Meyer, V. Fungal Biotechnology in Space: Why and How? in *Grand Challenges in Fungal Biotechnology* (ed. Nevalainen, H.) 501–535 (Springer International Publishing, Cham, 2020). doi:10.1007/978-3-030-29541-7_18.

46. Holliday, R. Early studies on recombination and DNA repair in *Ustilago maydis*. *DNA Repair* **3**, 671–682 (2004).

47. Cordero, R. J. B. *et al.* Radiation protection and structural stability of fungal melanin polylactic acid biocomposites in low Earth orbit. *Proceedings of the National Academy of Sciences* **122**, e2427118122 (2025).

48. Cordero, R. J. B., Dragotakes, Q., Friello, P. J. & Casadevall, A. Melanin protects Cryptococcus neoformans from spaceflight effects. *Environmental Microbiology Reports* **14**, 679–685 (2022).

49. Averesch, N. J. H., Shunk, G. K. & Kern, C. Cultivation of the Dematiaceous Fungus Cladosporium sphaerospermum Aboard the International Space Station and Effects of Ionizing Radiation. *Front. Microbiol.* **13**, (2022).

50. Koch, S. M. *et al.* Aspergillus niger as a cell factory for the production of pyomelanin, a molecule with UV-C radiation shielding activity. *Front. Microbiol.* **14**, (2023).

51. Brandić Lipińska, M. *et al.* Biological growth as an alternative approach to on and off-Earth construction. *Front. Built Environ.* **8**, (2022).

52. Acosta-Sampson, L. & King, J. Partially Folded Aggregation Intermediates of Human γD-, γC-, and γS-Crystallin Are Recognized and Bound by Human αB-Crystallin Chaperone. *Journal of Molecular Biology* **401**, 134–152 (2010).

53. Kosinski-Collins, M. S. & King, J. In vitro unfolding, refolding, and polymerization of human γD crystallin, a protein involved in cataract formation. *Protein Science* **12**, 480–490 (2003).

54. Schafheimer, N. & King, J. Tryptophan Cluster Protects Human γD-Crystallin from Ultraviolet Radiation-Induced Photoaggregation In Vitro. *Photochemistry and Photobiology* **89**, 1106–1115 (2013).

55. Lee, A. G., Mader, T. H., Gibson, C. R., Brunstetter, T. J. & Tarver, W. J. Space flight-associated neuro-ocular syndrome (SANS). *Eye* **32**, 1164–1167 (2018).

56. Lee, A. G. *et al.* Spaceflight associated neuro-ocular syndrome (SANS) and the neuro-ophthalmologic effects of microgravity: a review and an update. *npj Microgravity* **6**, 7 (2020).

57. Katsev, B. D. *et al.* Space Radiation Effects on the Glutathione Redox Cycle and Cataract Formation. *Aerospace Medicine and Human Performance* **97**, 354–361 (NaN/NaN/NaN).

58. Chylack, L. T. *et al.* NASA Study of Cataract in Astronauts (NASCA). Report 1: Cross-Sectional Study of the Relationship of Exposure to Space Radiation and Risk of Lens Opacity. *rare* **172**, 10–20 (2009).

59. McMackin, P., Adam, J., Griffin, S. & Hirsa, A. Amyloidogenesis via interfacial shear in a containerless biochemical reactor aboard the International Space Station. *npj Microgravity* **8**, 41 (2022).

60. Reichert, P. *et al.* Pembrolizumab microgravity crystallization experimentation. *npj Microgravity* **5**, 28 (2019).

61. Tovar, J. C. *et al.* Raspberry Pi–powered imaging for plant phenotyping. *Applications in Plant Sciences* **6**, e1031 (2018).

62. Bamsey, M. T., Paul, A.-L., Graham, T. & Ferl, R. J. Flexible imaging payload for real-time fluorescent biological imaging in parabolic, suborbital and space analog environments. *Life Sciences in Space Research* **3**, 32–44 (2014).

63. Sasidharan, K., Martinez-Vernon, A. S., Chen, J., Fu, T. & Soyer, O. S. A low-cost DIY device for high resolution, continuous measurement of microbial growth dynamics. Preprint at https://doi.org/10.1101/407742 (2018).

64. Deutzmann, J. S. *et al.* Low-Cost Clamp-On Photometers (ClampOD) and Tube Photometers (TubeOD) for Online Cell Density Determination. *Front. Microbiol.* **12**, (2022).

65. Barbé, B. *et al.* Pilot Testing of the “Turbidimeter”, a Simple, Universal Reader Intended to Complement and Enhance Bacterial Growth Detection in Manual Blood Culture Systems in Low-Resource Settings. *Diagnostics* **12**, (2022).

66. Case, N., Johnston, N. & Nadeau, J. Fluorescence Microscopy with Deep UV, Near UV, and Visible Excitation for In Situ Detection of Microorganisms. *Astrobiology* **24**, 300–317 (2024).

67. Schaefer, P. M., Kalinina, S., Rueck, A., von Arnim, C. A. F. & von Einem, B. NADH Autofluorescence—A Marker on its Way to Boost Bioenergetic Research. *Cytometry Part A* **95**, 34–46 (2019).

68. De Ruyck, J. *et al.* Towards the understanding of the absorption spectra of NAD(P)H/NAD(P)+ as a common indicator of dehydrogenase enzymatic activity. *Chemical Physics Letters* **450**, 119–122 (2007).

69. Wagnieres, G. A., Star, W. M. & Wilson, B. C. In Vivo Fluorescence Spectroscopy and Imaging for Oncological Applications. *Photochemistry and Photobiology* **68**, 603–632 (1998).

70. Pignataro, M. F., Herrera, M. G. & Dodero, V. I. Evaluation of Peptide/Protein Self-Assembly and Aggregation by Spectroscopic Methods. *Molecules* **25**, 4854 (2020).

71. Dougherty, M. *et al.* Results of the Micro-12 Flight Experiment: Effects of Microgravity on Shewanella oneidensis MR-.

72. Bellary, R. *et al.* Voltage, power, and energy production of a Shewanella oneidensis biofilm microbial fuel cell in microgravity. *J Emerg Invest* https://doi.org/10.59720/23-076 (2024) doi:10.59720/23-076.

73. Rowe, A. R., Chellamuthu, P., Lam, B., Okamoto, A. & Nealson, K. H. Marine sediments microbes capable of electrode oxidation as a surrogate for lithotrophic insoluble substrate metabolism. *Frontiers in Microbiology* **6**, 1–15 (2015).

74. Jangir, Y. *et al.* Isolation and Characterization of Electrochemically Active Subsurface Delftia and Azonexus Species. *Frontiers in Microbiology* **7**, (2016).

75. Jangir, Y. *et al.* In situ Electrochemical Studies of the Terrestrial Deep Subsurface Biosphere at the Sanford Underground Research Facility, South Dakota, USA. *Frontiers in Energy Research* **7**, (2019).

76. Harnisch, F., Deutzmann, J. S., Boto, S. T. & Rosenbaum, M. A. Microbial electrosynthesis: opportunities for microbial pure cultures. *Trends in Biotechnology* **42**, 1035–1047 (2024).

77. Li, Y. C. *et al.* An Easily Fabricated Low-Cost Potentiostat Coupled with User-Friendly Software for Introducing Students to Electrochemical Reactions and Electroanalytical Techniques. *J. Chem. Educ.* **95**, 1658–1661 (2018).

78. Elgrishi, N. *et al.* A Practical Beginner's Guide to Cyclic Voltammetry. *Journal of Chemical Education* acs.jchemed.7b00361 (2017) doi:10.1021/acs.jchemed.7b00361.

79. Guertin, S. Raspberry Pis for Space Guideline. *NASA WBS: 724297.40.49* http://nepp.nasa.gov/.

80. Mojica Decena, J., Wood, B., Martineau, R., Dennison, J. R. & Tayor, M. Radiation Damage Threshold of Satellite COTS Components: Raspberry Pi Zero for OPAL CubeSat. *Student Research Symposium* https://digitalcommons.usu.edu/mp_post/67 (2018).

81. Violette, D. P. Arduino-Raspberry Pi: Hobbyist Hardware and Radiation Total Dose Degradation. (2014).

82. Hanson, S. C. & Hayes, R. B. Radiation Hardness Assurance by Redundancy in Raspberry Pi Zero W Computation Metrics via Total Ionizing Dose 60Co Testing for Spacecraft Applications. *Health Physics* **128**, 457 (2025).

83. Ameur, Z. O. & Husein, M. M. Electrochemical Behavior of Potassium Ferricyanide in Aqueous and (w/o) Microemulsion Systems in the Presence of Dispersed Nickel Nanoparticles. *Separation Science and Technology* **48**, 681–689 (2013).

84. Harnisch, F. & Freguia, S. A basic tutorial on cyclic voltammetry for the investigation of electroactive microbial biofilms. *Chemistry - An Asian Journal* **7**, 466–475 (2012).

85. Gralnick, J. A. & Bond, D. R. Electron Transfer Beyond the Outer Membrane: Putting Electrons to Rest. *Annual Review of Microbiology* **77**, 517–539 (2023).

86. Lovley, D. R. Bug juice: harvesting electricity with microorganisms. *Nat Rev Microbiol* **4**, 497–508 (2006).

87. Juergensen, N., Weiler, J. R., Knoll, M. T., Gescher, J. & Edel, M. Strategic improvement of *Shewanella oneidensis* for biocatalysis: Approach to media refinement and scalable application in a microbial electrochemical system. *New Biotechnology* **85**, 31–38 (2025).

88. Yasri, N., Roberts, E. P. L. & Gunasekaran, S. The electrochemical perspective of bioelectrocatalytic activities in microbial electrolysis and microbial fuel cells. *Energy Reports* **5**, 1116–1136 (2019).

89. Xu, S., Jangir, Y. & El-Naggar, M. Y. Disentangling the roles of free and cytochrome-bound fl avins in extracellular electron transport from Shewanella oneidensis MR-1. *Electrochimica Acta* **198**, 49–55 (2016).

90. Gross, B. J. & El-Naggar, M. Y. A combined electrochemical and optical trapping platform for measuring single cell respiration rates at electrode interfaces. *Review of Scientific Instruments* **86**, 64301 (2015).

91. Torres, C. I. *et al.* A kinetic perspective on extracellular electron transfer by anode-respiring bacteria. *FEMS Microbiol Rev* **34**, 3–17 (2010).

92. Miller, S. D., Ford, K. C. & TerAvest, M. A. Outward and Inward Electron Transfer Occur through Distinct Interfacial Electron Transfer Mechanisms for Anaerobic Shewanella oneidensis. *ACS Omega* **11**, 70639–70648 (2026).

93. Lawless, C., Wilkinson, D. J., Young, A., Addinall, S. G. & Lydall, D. A. Colonyzer: automated quantification of micro-organism growth characteristics on solid agar. *BMC Bioinformatics* **11**, 287 (2010).

94. Moller, S., Kristensen, C. S., Poulsen, L. K., Carstensen, J. M. & Molin, S. Bacterial growth on surfaces: automated image analysis for quantification of growth rate-related parameters. *Applied and Environmental Microbiology* **61**, 741–748 (1995).